\documentclass[11pt,a4paper]{article}
\usepackage{jheppub}
\usepackage{tikz}
\usepackage{amsmath, amsfonts}
\usepackage{cleveref}
\usepackage{subcaption}
\usepackage{graphicx}
\usetikzlibrary{fit, positioning, calc}

\title{Pseudo-Evanescent Feynman Integrals from Local Subtraction}

\author[a]{Alessandro Georgoudis}
\author[b]{and Ben Page}

\affiliation[a]{Centre for Theoretical Physics, Department of Physics and Astronomy, Queen Mary University of London, Mile End Road, London E1 4NS, United Kingdom}
\affiliation[b]{Department of Physics and Astronomy, Ghent University, 9000 Ghent, Belgium}

\emailAdd{a.georgoudis@qmul.ac.uk}
\emailAdd{ben.page@ugent.be}

\abstract{
We introduce a new approach for the computation of the class of Feynman
integrals whose integrands vanish in strictly four-dimensions, so-called
``pseudo-evanescent'' integrals. We argue that, up to $\mathcal{O}(\epsilon)$
corrections, local subtraction techniques can be used to express
pseudo-evanescent integrals in terms of contributions from infrared and
ultraviolet regions of loop-momentum space.
We study two-loop examples and find that many pseudo-evanescent Feynman integrals are reduced
to either products of one-loop integrals or one-fold integrals thereof.
As a demonstration of the power of our approach, we use it to recompute the two-loop
all-plus five-point amplitude. We find that, up to scheme-dependent logarithms,
all contributions from soft and collinear regions cancel exactly against
known infrared structure and that the finite remainder is entirely given by
contributions from ultraviolet regions.
}

\preprint{QMUL-PH-25-07}

\newcommand{\eqnDiag}[1]{ \vcenter{\hbox{ #1}} }
\DeclareMathAlphabet\mathbfcal{OMS}{cmsy}{b}{n}

\newcommand{\eqnDiagr}[2][0pt]{%
  \vcenter{\hbox{\raisebox{#1}{#2}}}%
}

\begin{document}

\maketitle

\section{Introduction}

The computation of multi-leg scattering amplitudes in the Standard Model (SM)
beyond the one-loop level is a notoriously difficult problem.
Nevertheless, in recent years there has been a great deal of progress in pushing
forward the multiplicity frontier with cutting-edge two-loop amplitudes now
available for complicated multiscale processes such as $t\overline{t}j$~\cite{Badger:2025ljy, Badger:2024dxo} or
$Hb\overline{b}$~\cite{Badger:2024mir,Hartanto:2026xjz} production at the LHC.
This has been accompanied by similar progress in our ability to compute multi-leg Feynman integrals, for example, in computations of two-loop
six-point~\cite{Henn:2022ydo, Henn:2024ngj, Abreu:2024fei, Henn:2025xrc,Liu:2026hdp} and
three-loop five-point planar Feynman integrals~\cite{Liu:2024ont,Chicherin:2025mvc} for completely massless
processes.
An important technique employed in all such cutting-edge calculations is the method
of Feynman integral reduction, where the relations that Feynman integrals
satisfy are used to express amplitudes, or differential equations, as a linear
combination of a smaller set of integrals.
Importantly, this approach mandates the use of dimensional regularization, 
where the computations are performed in $D =
4-2\epsilon$ dimensions in order to consistently treat infrared and
ultraviolet divergences.
As the physics of interest is four dimensional, the physically relevant part of
the computation is extracted by carefully taking the $D \rightarrow
4$ limit.
Given the demanding nature of SM amplitude calculations, a natural question is,
therefore, to understand the impact of the four-dimensional limit on our
calculational framework.

It is well understood that careful understanding of the four-dimensional
limit can lead to simplifications in scattering amplitudes.
For example, in amplitudes with more than five external particles,
four-dimensionality of momentum space leads to further relations between Feynman
integrals, which has already played an important role in recent calculations of
two-loop six-point Feynman integrals~\cite{Henn:2022ydo, Henn:2024ngj,
Abreu:2024fei, Henn:2025xrc,Liu:2026hdp} and has been systematically classified at two loops in
refs.~\cite{Bargiela:2024rul,Bargiela:2025nqc}.
Moreover, Feynman integrals that are finite in the four-dimensional limit are
natural targets for Monte Carlo integration such as with
\texttt{pySecDec}~\cite{Borowka:2017idc,Heinrich:2021dbf} or tropical
approaches~\cite{Borinsky:2020rqs,Borinsky:2023jdv}.
This observation has motivated the development of a number of approaches for the
construction of ``quasi-finite''~\cite{vonManteuffel:2014qoa} and finite
bases~\cite{Agarwal:2020dye, Gambuti:2023eqh, delaCruz:2024xsm,  Dhani:2026cxx} of Feynman
integrals.
Interestingly, in ref.~\cite{DeAngelis:2025agn} bases of Feynman integrals and
associated algorithms were introduced whose master-integral coefficients are
free of poles in $\epsilon$.

In this work, we contribute to this four-dimensional program by considering
a particular class of Feynman integrals, known as ``pseudo-evanescent''
integrals (see e.g.~\cite{Gambuti:2023eqh}).
These integrals are a distinguished class, defined as those whose integrands vanish when evaluated on
four-dimensional loop momenta. Naively, such a vanishing integrand might lead
one to conclude that the integral would decouple in the four-dimensional limit
and therefore be ``evanescent''. However, subtle $\epsilon/\epsilon$
cancellations allow them to contribute, earning these integrals the name
of ``pseudo-evanescent integrals''.
At one-loop, pseudo-evanescent integrals play an important role in calculational
techniques. Specifically, they are a key ingredient of reduction techniques
where the integral coefficients are kept free of the dimensional regulator.
In practice, this comes at a cost of increasing the set of master
integrals to contain many pseudo-evanescent integrals.
Nevertheless, this difficulty is mitigated by the observation that one-loop
pseudo-evanescent integrals evaluate to rational functions of external
kinematics (see e.g.~\cite{Ossola:2006us}).

In contrast to the situation at the one-loop level, pseudo-evanescent integrals
and their advantages are less studied at two- and higher-loop orders.
Initial investigations include algorithmic techniques for classifying pseudo-evanescent integrals~\cite{Gambuti:2023eqh}.
Moreover, within the framework of off-shell methods of integrand construction, a
large amount of work has gone into understanding so-called rational terms of
ultraviolet origin~\cite{Pozzorini:2020hkx, Lang:2020nnl, Lang:2021hnw, Duhr:2023wsr}.
In this work, we contribute to this program by introducing an integration
approach tailored for pseudo-evanescent integrals at two loops that bypasses
standard master-integral-based techniques.
Specifically, following~\cite{Anastasiou:2018rib}, we develop a ``local subtraction'' method for the computation of pseudo-evanescent integrals. While local subtraction has already been successfully applied to many two-loop scattering processes~\cite{Anastasiou:2026kpm,Anastasiou:2025cvy,Anastasiou:2024xvk,Anastasiou:2022eym,Anastasiou:2020sdt}, we argue that it is also a valuable tool for the computation of pseudo-evanescent integrals.
To this end, we set up a computational scheme that allows us to compute pseudo-evanescent integrals up to $\mathcal{O}(\epsilon)$ corrections in a physically transparent way--as a sum of contributions from regions of momentum space associated to infrared and ultraviolet singularities.
Importantly, it turns out that the pseudo-evanescent integrals that we study 
in this work can be reduced to either products of one-loop integrals, or a one-fold
integral over a one-loop integral.
As a demonstration of the power of our techniques, we apply them to recompute
the two-loop, five-point all-plus gluonic helicity amplitude in pure
Yang-Mills theory~\cite{Badger:2013gxa, Gehrmann:2015bfy, Badger:2015lda,
Dunbar:2016aux, Badger:2019djh}.
Interestingly, such all-plus amplitudes have recently been shown to be connected
to Wilson loops with a Lagrangian
insertion~\cite{Chicherin:2022bov,Chicherin:2022zxo,Carrolo:2025pue,Chicherin:2025jej}
and amenable to chiral algebra bootstrap techniques~\cite{Morales:2025alm}.
An important feature of the five-point two-loop all-plus amplitude is that the
so-called ``finite remainder'', obtained after the subtraction of divergent
contributions, is given by a weight two transcendental
function~\cite{Gehrmann:2015bfy, Badger:2019djh}.
Similar features have also been observed in all-plus amplitudes for a self-dual
Higgs boson~\cite{Badger:2025uym}.
Our analysis makes this observation completely transparent, as we shall show,
the finite remainder is entirely controlled by the ultraviolet structure of the
integrand.

The paper is organized as follows: in section 2 we first introduce pseudo-evanescent
Feynman integrals and review the local counterterm technology of
ref.~\cite{Anastasiou:2018rib}. We then discuss how we apply local counterterms
to pseudo-evanescent integrals and how the application simplifies greatly in
this case. In section 3, we then apply our technology and undertake a
recomputation of the two-loop five-point all plus amplitude in pure Yang-Mills theory.
Finally, in section 4, we summarize and consider potential extensions of our
technology.

\section{Pseudo-Evanescent Integrals and Local Counterterms}
\label{sec:SubtractionMethodology}

The main object of study of this work is a class of
dimensionally-regulated integrals: pseudo-evanescent integrals.
In order to define this class of integral, we consider a general,
$l$-loop, dimensionally-regulated integral, taking the form
\begin{equation}
  I[f] = e^{l \epsilon \gamma_e} \int \prod_{j = 1}^l \frac{\mathrm{d}^D \ell_j}{i \pi^{D/2}} f(\ell_1, \ldots, \ell_l),
\end{equation}
where $\gamma_e$ is the Euler-Mascheroni constant.
Dimensional regularization comes in many flavours known as ``schemes'' (see
e.g.~\cite{Gnendiger:2017pys} for a summary). In a number of schemes the momenta of external particles are restricted to
be strictly four dimensional. This naturally singles out a four-dimensional
subspace of loop-momentum space, which has an important effect on the structure
of amplitude calculations.
To understand this, let us split the loop momenta into four-dimensional and
$(D-4)$-dimensional pieces.
Specifically, letting $v^{(4)}$ denote an arbitrary four-dimensional vector, we
write that
\begin{equation}
  \ell_i = \overline{\ell}_i + \tilde{\ell}_i, \qquad \text{such that} \qquad \overline{\ell}_i \cdot \tilde{\ell}_j = 0, \qquad \text{and} \qquad \tilde{\ell}_i \cdot v^{(4)} = 0.
\end{equation}
That is, we denote $D$-dimensional loop-momenta without any decoration,
four-dimensional loop-momenta with a bar, and 
$(D-4)$-dimensional loop-momenta with a tilde.
This decomposition now allows us to define pseudo-evanescent integrals.
Specifically, we define a pseudo-evanescent integral as an
integral of the form $I[f^{\text{pe}}]$ where $f^{\text{pe}}$ is some rational
function of loop momenta such that
\begin{equation}
 f^{\text{pe}}(\overline{\ell}_1, \ldots, \overline{\ell}_l) = 0.
\end{equation}
That is, the integrand of a pseudo-evanescent integral vanishes when evaluated on
four-dimensional loop-momentum configurations.

In this work, we will consider pseudo-evanescent Feynman integrals of the form
\begin{equation}
  I_\Gamma [N^{\text{pe}}(\ell_1, \ldots, \ell_l)] = I\left[\frac{N^{\text{pe}}(\ell_1, \ldots, \ell_l)}{\prod_{i \in \Gamma} D_i}  \right],
  \label{eq:peFeynmanIntegral}
\end{equation}
where $\Gamma$ is a graph, $N^{\text{pe}}$ is a polynomial in the loop momenta
and the $D_i$ are (inverse) Feynman propagators, each associated to an edge $i$
of the graph $\Gamma$. Frequently, we will represent the operator $I_\Gamma$ as the associated diagram $\Gamma$,
making use of the outgoing momentum convention.
As we consider pseudo-evanescent integrals, we impose that the integrand in
\cref{eq:peFeynmanIntegral} vanishes when evaluated on four-dimensional loop
momenta. As the denominator cannot vanish in this situation, we conclude that
the numerator must vanish on four-dimensional loop momenta, hence the label
$N^{\text{pe}}$.
Prominent examples of such integrals arise in amplitude computations where
$N^{\text{pe}}$ is explicitly Lorentz invariant in the $(D-4)$-dimensional
subspace. It is not hard to see that such numerators must be (polynomial) linear
combinations of the scalar product between the $(D-4)$-dimensional components of
loop momenta. We denote this as
\begin{equation}
  \mu_{ij} = \tilde{\ell}_i \cdot \tilde{\ell}_j.
\end{equation}
Clearly, these $\mu_{ij}$ vanish on four-dimensional configurations of loop
momenta as the individual $\tilde{\ell}_k$ are identically zero.
At the one-loop level, integrals of this form give rise to the so-called ``rational
piece'' of loop amplitudes.
As we will see, at higher loops, their analytic structure is much richer.

As integrands of pseudo-evanescent integrals vanish when evaluated on
four-dimensional momenta, one might intuit that the contribution of
pseudo-evanescent integrals to the scattering amplitude would be of
$\mathcal{O}(\epsilon)$ and therefore be unnecessary for four-dimensional
physics.
However, it is well known that this expectation is too naive, see e.g.
ref.~\cite{Weinzierl:2014iaa}.
In a very rough sense, the vanishing of the integrand in four dimensions may be
compensated by the fact that the integrand does not vanish fast enough in
singular regions.
In such examples, one heuristically expects that an integrand which vanishes in four
dimensions provides a factor of $\epsilon$, which can meet a pole in $\epsilon$ to give an
$\epsilon/\epsilon$ effect.
Indeed for one-loop Feynman integrals, it is well understood how to make this
intuition concrete, as pseudo-evanescent integrals can be handled by ``dimension
shifting'' relations~\cite{Bern:1993kr}.
Here, pseudo-evanescent integrals in $D$ dimensions can all be rewritten in
terms of higher-dimensional integrals that are not pseudo-evanescent, that is
\begin{equation}
  \int \frac{\mathrm{d}^D{\ell_1}}{i \pi^{D/2}} [\mu_{11}^s f(\ell_1)] = \frac{\Gamma([D+2s-4]/2)}{\Gamma([D-4]/2)} \int \frac{\mathrm{d}^{D+2s}{\ell_1}}{i \pi^{(D+2s)/2}} [f(\ell_1)].
  \label{eq:DimensionShifting}
\end{equation}
As the pre-factor on the right hand side of \cref{eq:DimensionShifting}
vanishes as $D\rightarrow 4$, we explicitly see that the pseudo-evanescent integral
only contributes in the limit if the dimension-shifted integral has a pole.
Moreover, in this limit, we need only compute the pole part of the
dimension-shifted integral, which is a much simpler calculation.

Unfortunately, while dimension-shifting identities extend to all loop orders~\cite{Tarasov:1996br,Tarasov:1997kx},
they cannot be used to reduce all pseudo-evanescent integrals to higher-dimensional
scalar integrals.
In this work, we take a different approach to the computation of
pseudo-evanescent integrals. Specifically, we employ the local counterterm
technology of ref.~\cite{Anastasiou:2018rib} and develop an approach to compute
pseudo-evanescent integrals up to $\mathcal{O}(\epsilon)$ corrections.
In the local counterterm approach, one makes use of the property that if the
integrand is ``locally convergent'',
then one can compute the four-dimensional limit of the integral by
simply integrating the four-dimensional limit of the integrand over a
four-dimensional phase space. Most Feynman integrals are not locally convergent,
and hence it is necessary to construct local counterterms in order to be able
to take the four-dimensional limit at the integrand level.
In ref.~\cite{Anastasiou:2018rib}, a technology was introduced 
to construct such local counterterms for many two-loop Feynman integrals.
Specifically, for a two-loop Feynman integrand $f(\ell_1,
\ell_2)$, the approach is to construct a local counterterm expression
$f_{\mathrm{approx}}$ such that
\begin{equation}
  \int d^D \ell_1 d^D \ell_2 \left[ f(\ell_1, \ell_2)  - f_{\mathrm{approx}}(\ell_1, \ell_2) \right] =
  \int d^4 \overline{\ell}_1 d^4 \overline{\ell}_2 \left[ f(\overline{\ell}_1, \overline{\ell}_2)  - f_{\mathrm{approx}}(\overline{\ell}_1, \overline{\ell}_2) \right] + \mathcal{O}(\epsilon).
  \label{eq:LocalCountertermConvergence}
\end{equation}
This statement is especially appealing in the context of pseudo-evanescent
integrals, as their integrands vanish in the four-dimensional limit. If we are
able to construct counterterms which also share this property, we see that the
right hand side of \cref{eq:LocalCountertermConvergence} becomes
$\mathcal{O}(\epsilon)$. For appropriately constructed counterterms, we
therefore are led to the foundational observation of this work,
\begin{equation}
  \int d^D \ell_1 d^D \ell_2 \left[ f^{\text{pe}}(\ell_1, \ell_2)   \right] =
  \int d^D \ell_1 d^D \ell_2 \left[ f^{\text{pe}}_{\mathrm{approx}}(\ell_1, \ell_2) \right] + \mathcal{O}(\epsilon).
  \label{eq:MuCountertermEquivalence}
\end{equation}
This simple correspondence, between pseudo-evanescent integrals and their local
counterterms, is very powerful. We will see that it leads to a greatly
simplified computational strategy for pseudo-evanescent integrals, as well as a
physically transparent description.
In the rest of this section, we will explore how to systematically construct
appropriate counterterms for two-loop pseudo-evanescent integrals that allow us
to make use of \cref{eq:MuCountertermEquivalence}.

\subsection{Review of Singular Regions and Power-Counting}
\label{sec:SingularRegions}

In order to ascertain if an integral is locally convergent, the standard approach
is to test how the integrand and integration measure scale as the
loop momenta approach a pinch surface~\cite{Landau:1959fi, Weinberg:1959nj,
Sterman:1978bi}.
Here, we review the basics of such an analysis, following
ref.~\cite{Anastasiou:2018rib}. The regions are associated to configurations of loop
momenta where internal line momenta become either soft,
collinear to an external momentum (infrared singularities) or large (an
ultraviolet singularity). In order to detect if a Feynman integral is locally
convergent in such regions, one employs a power-counting approach.
In the following, we make use of a reference hard scale $Q$, and a scaling
parameter $\lambda$. In any given region, the combined integrand and integration measure
scale as $\lambda^p$.
Generically, we will consider $\lambda$ to be small such that the power-counting
analysis warns us of a singular region when $p \le 0$. If $p=0$, this is known
as a logarithmic divergence.
If $p<0$, this is known as a power divergence. Such power divergences are
categorized by their strength. For example, we call $p=-1$ a linear divergence.

We first consider the regions associated to infrared singularities. In a soft
region, the momenta of some set of internal lines approach zero. Labeling the
momenta of such a line as $\ell$, the components of $\ell$ and the associated
integration measure scale as
\begin{equation}
  \label{eq:SoftScaling}
   \ell^\mu \sim \lambda Q, \qquad \qquad 
  \mathrm{d}^D \ell \sim \lambda^D.
\end{equation}
Therefore, for $s$ independent soft loop momenta, the measure scales as
$\lambda^{4s}$ in 4 dimensions.
In a collinear region, the momenta of a set of internal lines become
collinear to that of a light-like external momentum $p_j$. Labeling such a
momentum as $\ell$, we parameterize this approach as
\begin{equation}
  \label{eq:CollinearParameterization}
  \ell = x_j p_j + \beta_j \eta_j + \ell_{\bot},
\end{equation}
where $\eta_j$ is a light-like reference vector (e.g. the parity conjugate of
$p_j$), and $\ell_{\bot}$ is orthogonal to both $p_j$ and $\eta_j$.
We remark that, in the collinear limit, $x_j$ becomes the collinearity fraction and
can be computed via
\begin{equation}
  \label{eq:CollinearityDefinition}
  x_j = \frac{\ell \cdot \eta_j}{p_j \cdot \eta_j}.
\end{equation}
In the parameterization of \cref{eq:CollinearParameterization}, the momenta and
the associated integration measure in the collinear region scale as
\begin{equation}
  \label{eq:CollinearScaling}
  \beta_j \sim \lambda,
  \qquad \qquad
  \ell_{\bot}^\mu \sim \sqrt{\lambda},
  \qquad \qquad
  \mathrm{d}^D \ell \sim \lambda^{D/2}.
\end{equation}
Therefore, for $c$ independent collinear loop momenta, in strictly
four-dimensions, the measure scales as $\lambda^{2c}$.

Next, we consider ultraviolet regions, where some set of loop momenta
become large. In order to describe these regions, we will work graphically.
For any Feynman integral, we can associate a graph, consisting of the loop-momentum
dependent edges of the integral.
Such a graph contains (potentially non-proper) closed-loop subgraphs. We denote
each such closed-loop subgraph as $\Gamma_i^{(l)}$, where $l$ is the loop order of the
subgraph.
To any subgraph $\Gamma_i^{(l)}$, we associate an ultraviolet region where the momenta in the subgraph scales as
\begin{equation}
    k_j \sim \frac{1}{\lambda} Q \qquad \text{for all} \qquad k_j \in \Gamma_i^{(l)}.
\end{equation}
Denoting the independent loop momenta in $\Gamma_i^{(l)}$ as $\ell_1, \ldots,
\ell_l$, the measure scales as
\begin{equation}
  \prod_{i=1}^l \mathrm{d}^D\ell_i \sim \lambda^{-D l}.
\end{equation}
That is, in strictly four dimensions, the measure scales under such a limit as
$\lambda^{-4l}$ .
If an $L$-loop Feynman integral is power-counting divergent in a region
associated to a proper, $l<L$, subgraph $\Gamma_i^{(l)}$ we call such a
divergence a ``sub-divergence''. If the $L$-loop integral is power-counting
divergent in a region associated to the full graph we call
this a global divergence.

\subsection{Power Counting and Pseudo-Evanescent Integrals}
\label{sec:EvanescentPowerCounting}

Having discussed power-counting rules to determine if a generic Feynman
integral fails to be locally convergent, let us now consider applying them to
pseudo-evanescent Feynman integrals.
As we will see, in a number of regions pseudo-evanescent integrals exhibit
improved power counting, which will greatly simplify their calculation via the
counterterm strategy.
We consider integrals of the form
$I_{\Gamma}[N^{\text{pe}}]$ for some graph $\Gamma$, and some numerator polynomial $N^{\text{pe}}$,
where $N^{\text{pe}}$ vanishes when evaluated on four-dimensional loop momenta,
and, hence, is
a (polynomial-)linear combination of $\mu_{ij}$. 
As such, we can break the problem of determining the power counting down into
two pieces: understanding the power-counting contributions from the $\mu_{ij}$
in the numerator,
and understanding the power-counting contributions arising from the denominators
of $\Gamma$ and the integration measure, or, equivalently, that of the associated
scalar integral.

\paragraph{Pseudo-Evanescent Power Counting at One Loop}
As a warmup, we begin with one-loop pseudo-evanescent integrals.
Here, there is only one $\epsilon$-dimensional scalar product,
$\mu_{11}$. As such, the numerator of all pseudo-evanescent integrals under
consideration must be proportional to $\mu_{11}$.
Let us consider how this behaves in the various singular regions.
First, we observe that $\mu_{11}$ is invariant under shifts of the loop
momentum by any external momentum. This allows us to always consider the
singular momentum to be $\ell_1$.
In a one-loop soft region, by \cref{eq:SoftScaling}, we therefore have that
\begin{equation}
  \mu_{11} \sim \lambda^2.
\end{equation}
By similar logic and \cref{eq:CollinearScaling}, in a one-loop collinear region we have that
\begin{equation}
  \mu_{11} \sim \lambda.
\end{equation}
We therefore see that such pseudo-evanescent integrals exhibit improved
infrared power counting with respect to the associated scalar integral.
The final region to consider is the ultraviolet, where we have that
\begin{equation}
   \mu_{11} \sim \frac{1}{\lambda^2}.
   \label{eq:OneLoopMuUVPowerCounting}
\end{equation}
As such, factors of $\mu_{11}$ in the numerator worsen ultraviolet
power-counting behavior.

\begin{figure}

  \begin{subfigure}{0.45\textwidth}
  \centering
    $\eqnDiag{\includegraphics[scale=0.5]{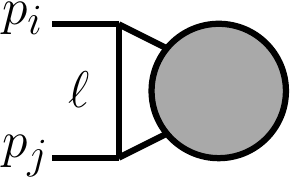}}$,
  \caption{}
  \label{fig:SoftPropagators}
  \end{subfigure}
  \begin{subfigure}{0.45\textwidth}
  \centering
    $\eqnDiag{\includegraphics[scale=0.5]{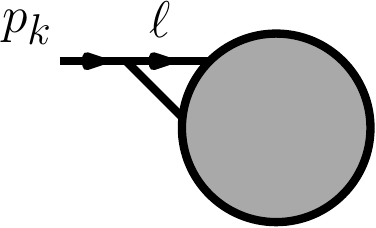}}$,
  \caption{}
  \label{fig:CollinearPropagators}
  \end{subfigure}
  \caption{Sub-diagrams which give rise to logarithmic scaling behavior in
    singly singular limits.
    The gray blob represents some general sub-diagram. The momenta $p_i, p_j$ and
$p_k$ are on-shell. The momentum $p_k$ is light-like.
  }

  \label{fig:SinglySingularSubdiagrams}

\end{figure}
Let us now consider the power-counting contributions associated to the scalar integrals
themselves.
We consider scalar integrals that arise in Feynman-gauge scattering
amplitudes. As such, no propagators are doubled.
It is well understood that, at one loop, there are only two cases where such scalar integrals exhibit
divergent scaling behavior in infrared regions. These two cases are
logarithmically divergent and arise from particular sets of propagators. These
can be understood diagrammatically and are depicted in
\cref{fig:SinglySingularSubdiagrams}. For simplicity, we consider integrals
arising in a massless theory, though the logic is easily extended.
The logarithmically-divergent single-soft case arises when the loop momentum
in a massless line that is between two on-shell external particles approaches zero. We depict the relevant set of propagators in
\cref{fig:SoftPropagators}, reducing the remaining part of the diagram to a
blob. In the strict soft limit, the loop momentum $\ell$ of
\cref{fig:SoftPropagators} is zero.
In the logarithmically-divergent collinear case, the loop momenta in two
sequential massless edges become proportional to the external lightlike
momentum $p_k$ that meets them.
We depict this subset of propagators in \cref{fig:CollinearPropagators}, again
reducing the remaining part of the diagram to a blob.
In the strict collinear limit, the loop momentum $\ell$ in
\cref{fig:CollinearPropagators} is proportional to $p_k$.
Given that numerators of (pseudo-)evanescent integrals must vanish in such regions, then we
conclude that all relevant (pseudo-)evanescent integrals are locally convergent in
the soft and collinear region.
In contrast, if we consider the ultraviolet region, it is clear, by
\cref{eq:OneLoopMuUVPowerCounting}, the ultraviolet power-counting can be made
arbitrarily large by introducing factors of $\mu_{11}$.
We are therefore able to generically conclude that the only relevant singular
region for Feynman integrals relevant for fixed-angle scattering amplitudes at
one-loop is the ultraviolet region.

Let us now apply this logic to a concrete example. We consider the family of box
integrals
\begin{equation}
  I_{\text{box}}^{(r)} = \eqnDiag{\includegraphics[scale=0.6]{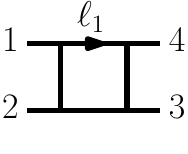}}[\mu_{11}^r],
\end{equation}
where $r$ is some non-negative integer.
For all $r>0$, $I_{\text{box}}^{(r)}$ exhibits convergent scaling in the
infrared regions as the logarithmic scaling of the scalar box is suppressed by a
factor of $\mu_{11}$ in the numerator.
Nevertheless, in the ultraviolet region, the divergence behavior depends on $r$.
Specifically, $I_{\text{box}}^{(1)}$ has ultraviolet quadratically convergent
scaling, while $I_{\text{box}}^{(2)}$ and $I_{\text{box}}^{(3)}$ have
ultraviolet logarithmic and quadratically divergent scaling respectively.

\paragraph{(Pseudo-)Evanescent Power Counting at Two Loops}

Let us now consider two-loop pseudo-evanescent Feynman integrals, making a
number of general observations on the power-counting behavior of
pseudo-evanescent integrals at two loops.
In contrast to the one-loop case, there are three possible
$(D-4)$-dimensional scalar products: $\mu_{11}$, $\mu_{22}$ and $\mu_{12}$. This renders
the analysis more intricate.
Moreover, when considering singular regions, there are a larger number of
configurations to consider. Specifically, each loop momentum can independently
be soft, collinear, ultraviolet or not enter a singular region. We refer to this
non-singular case as the loop momentum being ``hard''.
We refer to the regions where one or both loop momenta are singular as
``single'' and ``double'' regions respectively.

We begin by understanding the behavior of the $\mu_{ij}$ in the double singular
regions. Similar to the one-loop case, it is easy to see that any factor of
$\mu_{ij}$ will worsen the ultraviolet powercounting. This is true either if one
or both loop momenta are ultraviolet. Therefore, we see that the ultraviolet
behavior must always be considered case by case.
We are thus reduced to the analysis of the double-infrared regions, where
factors of $\mu_{ij}$ in the numerator will again give rise to a suppression.
There are naturally three classes of double-infrared regions: double-soft,
soft-collinear and double-collinear.
As in the one-loop case, we exploit that the $\mu_{ij}$ are invariant under
shifts of the loop momentum by external momenta, and so we always assume that
$\ell_1$ and $\ell_2$ are the singular momenta.
In any double-soft region, we apply \cref{eq:SoftScaling} for both loop momenta,
finding that
\begin{equation}
  \mu_{11} \sim \mu_{22} \sim \mu_{12} \sim \lambda^2.
  \label{eq:MuijDoubleSoftScaling}
\end{equation}
Importantly, we see that each $\mu_{ij}$ has the same scaling in the double-soft
region, each offering a quadratic suppression.
Similarly, in a double-collinear region, we apply \cref{eq:CollinearScaling} for
both loop momenta, finding
\begin{equation}
  \mu_{11} \sim \mu_{22} \sim \mu_{12} \sim \lambda.
  \label{eq:MuijDoubleCollinearScaling}
\end{equation}
Once again, we see that each $\mu_{ij}$ has the same scaling in a
double-collinear region, now offering a linear suppression.
Interestingly, this democratic scaling behavior between the $\mu_{ij}$ does not
hold if we consider the soft-collinear region. Without loss of
generality, we consider a region where $\ell_1$ goes soft and $\ell_2$ goes
collinear to some external momentum. Applying \cref{eq:SoftScaling} for $\ell_1$
and \cref{eq:CollinearScaling} $\ell_2$ leads to a scaling behavior of
\begin{equation}
  \mu_{11} \sim \lambda^2, \quad \mu_{22} \sim \lambda, \quad \mu_{12} \sim \lambda^{3/2}.
  \label{eq:MuijSoftCollinearScaling}
\end{equation}
We therefore see that the soft-collinear scaling behavior of pseudo-evanescent integrals with
$\mu_{ij}$ factors in the numerator depends on the particular $\mu_{ij}$ under consideration.

\begin{figure}[t]
  \begin{subfigure}{0.3\textwidth}
  \centering
  $\eqnDiag{\includegraphics[scale=0.5]{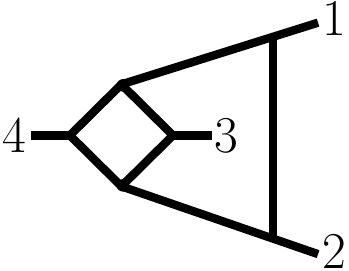}}$
  \caption{}
  \label{fig:NPDoubleBox}
  \end{subfigure}
  \begin{subfigure}{0.3\textwidth}
  \centering
  $\eqnDiag{\includegraphics[scale=0.6]{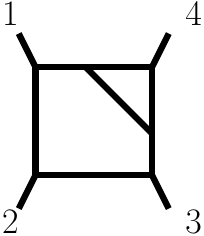}}$
  \caption{}
  \label{fig:beetle}
  \end{subfigure}
  \begin{subfigure}{0.3\textwidth}
  \centering
  $\eqnDiag{\includegraphics[scale=0.6]{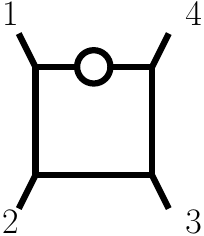}}$
  \caption{}
  \label{fig:selfEnergy}
  \end{subfigure}
  \caption{Example two-loop topologies that exhibit power-like scaling behavior in their respective
    scalar integrals. The scalar non-planar double box integral of
    \cref{fig:NPDoubleBox} exhibits a linear double-soft singularity, while the
    scalar beetle diagram of \cref{fig:beetle} exhibits linear double-soft,
    soft-collinear and double-collinear singularities.
    The self-energy diagram of \cref{fig:selfEnergy} exhibits, for example, a quadratic
    single-soft singularity.
  }
  \label{fig:PowerDivergentTopologies}
\end{figure}

Next, we consider the possible scaling behaviors of scalar integrals in
double-infrared regions at two loops.
Broadly speaking, most two-loop scalar Feynman integrals without double
propagators are at worst logarithmically divergent all double-infrared
regions.
However, it is well known (see
e.g. \cite{Gambuti:2023eqh}) that some two-loop Feynman integral topologies correspond to
scalar integrals with power-divergence behavior in some double-infrared regions.
Examples of double-infrared power-divergent topologies can be found in
\cref{fig:PowerDivergentTopologies}.
In \cref{fig:NPDoubleBox}, there are two double-soft regions where the scalar
integral is linearly divergent.
However, looking to the double-soft scaling of \cref{eq:MuijDoubleSoftScaling},
any factor of $\mu_{ij}$ provides a suppression, and such
pseduo-evanescent integrands are locally convergent in this region.
However, if we consider \cref{fig:beetle}, we find that it exhibits a linear
power divergence in double-collinear and soft-collinear regions. Looking to
\cref{eq:MuijDoubleCollinearScaling} and \cref{eq:MuijSoftCollinearScaling}, we
see that in such cases factors of $\mu_{ij}$ are insufficient to render the
integral double-infrared safe.
Similar conclusions can be drawn about the self-energy diagram in \cref{fig:selfEnergy}.
For simplicity, and as it is sufficient for the applications to two-loop
all-plus amplitudes that we study in this work, we will restrict our study to
pseudo-evanescent integrands that are convergent in double-infrared regions,
leaving more intricate cases such as the diagrams of \cref{fig:beetle} and
\cref{fig:selfEnergy} to future work.

\begin{figure}[t]
\centering
\begin{tikzpicture}[
    every node/.style={inner sep=0pt},
    lbl/.style={font=\small, align=center}
]

\node (DB) at (0,0) {\includegraphics[scale=1.0]{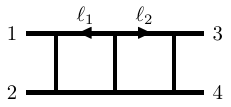}};
\node[right=-1pt of DB, yshift=-5pt] {$\big[\mu_{22}^2\big]$};

\node (UV)   at ($(DB.east)+(4, 1.1)$)  {\includegraphics[scale=0.8]{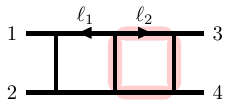}};
\node[lbl, below=3pt of UV]   {Ultraviolet};

\node (Soft) at ($(DB.east)+(7.5, 1.1)$)  {\includegraphics[scale=0.8]{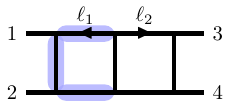}};
\node[lbl, below=3pt of Soft] {Soft};

\node (C1)   at ($(DB.east)+(4,-1.1)$)  {\includegraphics[scale=0.8]{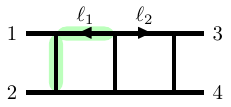}};
\node[lbl, below=3pt of C1]   {Collinear to $p_1$};

\node (C2)   at ($(DB.east)+(7.5,-1.1)$)  {\includegraphics[scale=0.8]{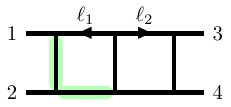}};
\node[lbl, below=3pt of C2]   {Collinear to $p_2$};

\draw[->, >=stealth, thick] ($(DB.east)+(1.2,-0.2)$) -- ($(DB.east)+(2.2,-0.2)$);

\end{tikzpicture}
\caption{In this figure we illustrate the different regions contributing to the finite part of the diagram with a $\mu_{22}$ insertion. 
For the ultraviolet region, we highlight the propagators in the $\ell_2$ loop that generate the divergence. 
In the language of counterterms, the red subgraph corresponds to the $\Gamma^{(1)}$ appearing in \cref{eq:UltravioletCounterterm}. 
The blue lines denote the propagators contributing to the soft region, arising when the propagator connecting legs $p_1$ and $p_2$ becomes soft. 
In this configuration, the shrinking propagator is the only one not highlighted in the $\ell_1$ loop. 
Finally, the green lines indicate the propagators that must be integrated out according to the collinear counterterm construction discussed in section \ref{sec:SingleIRCounterterms}.}
\label{fig:regionsdivdb}
\end{figure}
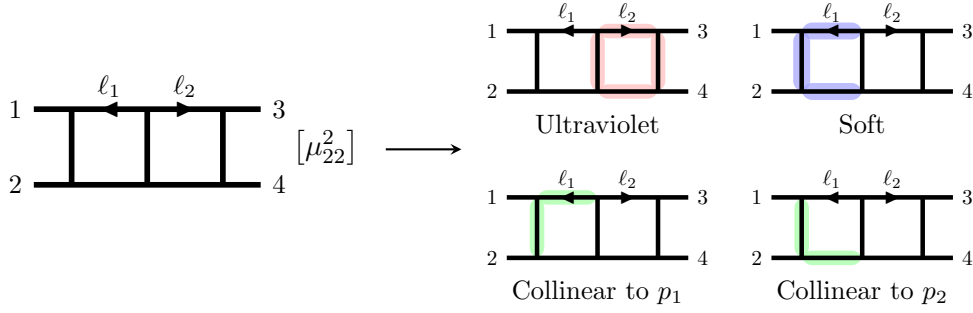

Having considered the effects of double-singular regions, it remains to consider
power-counting behavior in single-singular regions. Naturally, as one of the two loop
momenta remains hard, the scaling of the $\mu_{ij}$ will not be democratic.
Without loss of generality, we will consider $\ell_1$ to be a singular momentum,
while $\ell_2$ will remain hard.
As in the one-loop analysis, it is clear that factors of $\mu_{ij}$ can
arbitrarily worsen ultraviolet power-counting. Therefore, we cannot a priori avoid
analyzing any integral topologies.
Thus, it remains to consider the infrared regions. In a single soft region,
we have that
\begin{equation}
  \mu_{11} \sim \lambda^2, \qquad \mu_{12} \sim \lambda, \qquad \mu_{22} \sim 1.
\end{equation}
While in a single collinear region, we have that 
\begin{equation}
  \mu_{11} \sim \lambda, \qquad \mu_{12} \sim \sqrt{\lambda}, \qquad \mu_{22} \sim 1.
\end{equation}
Naturally, we see that $\mu_{11}$ scales identically as it did at one loop, but
factors of $\mu_{22}$ do not vanish. The mixed case, $\mu_{12}$, still vanishes
in each region, though more slowly. Therefore, in a single-infrared region
associated to the loop momentum $\ell_k$, a
factor of $\mu_{ij}$ only provides a suppression if at least one of  $i = k$ or
$j = k$ holds.
The scaling behavior of the associated scalar integrals is also comparatively
simple. Here, the diagrammatic configurations of concern are essentially
one-loop in nature and greatly simplified. That is, all lines in the sub-loop
associated to the non-singular momentum scale as $\mathcal{O}(1)$.
In almost all cases, this results in one of the two one-loop diagrams in
\cref{fig:SinglySingularSubdiagrams}. However, self-energy diagrams, such as
those depicted in \cref{fig:selfEnergy} are more subtle.
Specifically, if we consider a region where the loop momentum associated to
self-energy diagram is hard, the
resulting one-loop diagram now contains a doubled propagator. 
As such, the
scalar self-energy diagram exhibits power-like singular behavior in both single-soft
and single-collinear limits. Nevertheless, we have already excluded this diagram
from our analysis as it does not arise in the all-plus amplitudes we study.
To close, we consider a concrete example of a two-loop pseudo-evanescent Feynman
integral: the double box with a factor of $\mu_{22}$ in the numerator (see \cref{fig:regionsdivdb}).
This example exhibits divergent power counting  in single soft, collinear and ultraviolet regions.
Specifically, we see that the integrand does not vanish if the loop momentum
enters the soft region associated to legs $1$ and $2$, or either of the
associated collinear regions. Also the integrand exhibits an ultraviolet divergent region as $\ell_2$ becomes large. 

\subsection{Counterterms and Two-Loop Pseudo-Evanescent Integrals}
\label{sec:CountertermConstruction}

In this section, we develop a local counterterm approach to render a
pseudo-evanescent Feynman integral locally convergent in each region. Our counterterms will themselves be pseudo-evanescent Feynman integrals, and so by \cref{eq:MuCountertermEquivalence}, this allows us to calculate them up to $\mathcal{O}(\epsilon)$ corrections.
Given the discussion of the previous section, we focus on the broad class of
two-loop pseudo-evanescent Feynman integrals $I_\Gamma[N^{\text{pe}}]$ which
exhibit convergent power-counting in double-infrared regions.
Our task is to systematically construct local counterterms such that the
subtracted integral has convergent power counting in all singular regions.
Practically, this is made easier by the relatively mild behavior of
pseudo-evanescent Feynman integrals in singular regions. As we will see, this
allows us to avoid a potentially large proliferation of counterterms.
To this end, we first discuss how to construct counterterms that improve the
power-counting behavior in individual singular regions, both infrared and
ultraviolet.
Thereafter, we discuss how to apply this counterterm construction to render a
pseudo-evanescent Feynman integral convergent in all singular regions
simultaneously.

\subsubsection{Single-Infrared Counterterms}
\label{sec:SingleIRCounterterms}

Let us begin with constructing counterterms relevant for the single-soft region. 
Our approach is that of ref.~\cite{Anastasiou:2018rib}.
As discussed earlier, we restrict to two-loop Feynman integrals that do not
contain self-energy insertions.
We recall that, for such a Feynman integral to develop divergent power-counting
in this region, it must contain the sub-diagram depicted in
\cref{fig:SoftPropagators}. As this occurs when the line momentum
between particles $i$ and $j$ goes soft, we call this the $(i, j)$ soft region.
Moreover, as we have discussed, the two-loop integrals under
consideration are at worst logarithmically singular. We therefore require to
only match the integrand to leading power in the limit.
To this end, we define an approximation for the $(i,j)$ soft region as
\begin{equation}
  \mathcal{S}_{(i,j)} \left[ \eqnDiag{\includegraphics[scale=0.5]{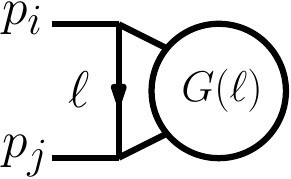}} \right] =
\eqnDiag{\includegraphics[scale=0.5]{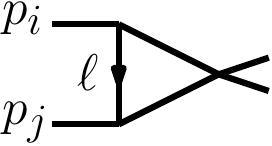}}[1] \,\, G(0).
\label{eq:LogarithmicSoftCounterterm}
\end{equation}
Here, we single out the three propagators that are on shell in the $(i,j)$ soft
limit, and denote the remaining piece of the integral as $G(\ell)$. That is,
$G(\ell)$ is a lower loop integral, to which we consider $\ell$ as an external
momentum.
On the right hand side of \cref{eq:LogarithmicSoftCounterterm}, the associated
triangle integral matches the singular denominators and the $G(0)$ factor
provides the correct weight in the limit.
The operator $\mathcal{S}_{(i,j)}$ can naturally be extended to be a linear
operator by defining its action on an integral with convergent power counting in
the $(i,j)$ soft region to be zero.
An important property of the definition of \cref{eq:LogarithmicSoftCounterterm}
is that, when applied to two-loop integrals, the counterterm is a product of
one-loop integrals, and therefore significantly easier to calculate.

Let us next consider constructing counterterms for (pseudo-)evanescent Feynman
integrals that are divergent in the single-collinear region. We again restrict
to two-loop Feynman integrals that do not contain self-energy insertions. As
such, for a Feynman integral to develop divergent power counting, it must
contain the subdiagram depicted in \cref{fig:CollinearPropagators}.
Moreover, it can be at worst logarithmically singular in this region, and we
must only construct a counterterm that matches to leading power.
The act of making an appropriate approximation in the collinear region is more
subtle, and is one of the innovations made in ref.~\cite{Anastasiou:2018rib}. As
the collinear region is not a point in momentum space like the soft region, but
in fact a line, we expect an integral over all collinear configurations.
Following the discussion of ref.~\cite{Anastasiou:2018rib} we define an
approximation to our target Feynman integrals in the collinear region through
\begin{equation}
    \mathcal{C}_{j}
    \left[\! \eqnDiag{\includegraphics[scale=0.35]{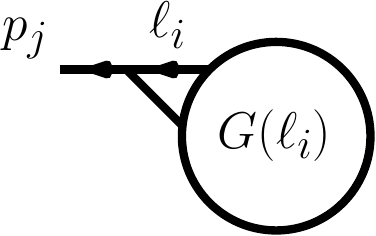}}
      \, \right] \!=\!
    e^{\epsilon \gamma_E} \! \int \! \frac{\mathrm{d}^D\ell_i}{i \pi^{D/2}} \left[ \frac{1}{\ell_i^2 (\ell_i \!-\! p_j)^2}  - \frac{1}{\left( \ell_i^2 \!-\! M^2 \right)\left( [\ell_i\!-\!p_j]^2 \!-\! M^2 \right)}\right] G(x_j[\ell_i] p_j),
\label{eq:CollinearLocalCounterterm}
\end{equation}
where $x_j[\ell_i]$ is as defined in \cref{eq:CollinearityDefinition} and $M$ is an
auxiliary mass scale.
On the left-hand side of \cref{eq:CollinearLocalCounterterm}, we expose the two
propagators which go on-shell in the collinear limit, while the function
$G(\ell)$ encompasses all of the remaining parts of the integral, both
propagators and numerator. Similar to the soft case, $G(\ell)$ is a
one-lower-loop integral, to which $\ell$ is regarded as an external momentum.
A crucial benefit of the counterterm definition in
\cref{eq:CollinearLocalCounterterm} is that the integral over $\ell_i$ is easy
to perform, as first done in ref.~\cite{Anastasiou:2018rib}. The result is that
\begin{equation}
    \mathcal{C}_{j} \left[\eqnDiag{\includegraphics[scale=0.4]{graphics/collinearHighlight.pdf}} \,\, \right] =
    -\frac{\Gamma(1+\epsilon) M^{-2\epsilon}}{\epsilon} \int_0^1 \mathrm{d} x \, G(x p_j),
\label{eq:CollinearCounterterm}
\end{equation}
where we manifestly see a pole in $\epsilon$ corresponding to the collinear singularity.
For the benefit of the reader, we record a detailed derivation of
\cref{eq:CollinearCounterterm} in \cref{app:CollinearKernelDerivation}. We note
that the right-hand-side of \cref{eq:CollinearCounterterm} contains a number of
factors and for later purposes we define
\begin{equation}
  \label{eq:collienearSsub}
   \overline{\mathcal{C}}_j\left[ I \right] = - \frac{M^{2 \epsilon}}{\Gamma(1+\epsilon)} \mathcal{C}_j\left[ I \right],
\end{equation}
in order to avoid the cluttering of notation.

Let us consider the structure of the collinear counterterm definition in
\cref{eq:CollinearLocalCounterterm}. On the right-hand side, the first term
effectively localizes the $G(\ell)$ piece of the integral to the singular
configuration. However, in doing so, the integration involving the two collinear
propagators becomes ultraviolet divergent.
In order to avoid introducing a spurious ultraviolet divergence, one includes
the second term, which is regular on the collinear pinch surface but shares the
same ultraviolet behavior.
Altogether, we see that the collinear counterterm matches a Feynman integral
to leading power in the collinear region, without introducing any extra
divergences.
Importantly, when applied to two-loop integrals, the collinear counterterm is
given by a one-fold integral over a one-loop integral. Naturally, this is a
significantly simpler calculation to perform.
Finally, we close the discussion of the collinear counterterm by noting that, to
render the second term in \cref{eq:CollinearCounterterm} finite in the collinear
region, we have introduced an auxiliary mass scale $M$.
When applying this collinear counterterm to an arbitrary Feynman integral, the
result develops a non-trivial dependence on $M$.
Nevertheless, when applying our formalism to pseudo-evanescent Feynman
integrals, we see that the counterterms and the original integral must match up
to $\mathcal{O}(\epsilon)$ corrections. We therefore conclude that the
dependence on $M$ will only arise in $\mathcal{O}(\epsilon)$ corrections.
Nevertheless, such cancellations may prove intricate.

\subsubsection{Ultraviolet Counterterms}

Let us now discuss counterterms for ultraviolet regions. 
In contrast to our infrared discussion, we expect counterterms in ultraviolet
cases to be more intricate than those for infrared regions as we consider cases
where the power counting in the ultraviolet can be power like.
Therefore, it is necessary to define counterterms which match the integrand to
higher orders in the ultraviolet.
Our approach to construct ultraviolet counterterms is based upon Taylor
expansion around the ultraviolet region.
This technique is well known~\cite{Misiak:1994zw,Chetyrkin:1997fm}, and is
equivalent to that used in
refs.~\cite{Pozzorini:2020hkx,Lang:2020nnl,Lang:2021hnw}. Amusingly, the
approach has been re-discovered many times, for example in the
``four-dimensional regularization/renormalization''~\cite{Pittau:2012zd}.

To begin, we consider an ultraviolet region associated to an $l$-loop subdiagram
$\Gamma^{(l)}$ of an $L$-loop graph $\Gamma$.
We denote the integrand associated to the $L$-loop graph $\Gamma$ as
$G(\ell_1, \cdots, \ell_L)$ and the loop momentum associated to
$\Gamma^{(l)}$ as $\ell_1, \ldots, \ell_l$.
The important structural observation is that the Taylor expansion of the
integrand around the ultraviolet limit associated to $\Gamma^{(l)}$ can be
written as a sum of terms which factorize. In full generality, we can write
\begin{equation}
  \lambda^{-4 l} G (\lambda^{-1} \ell_1, \ldots, \lambda^{-1}\ell_l, \ldots, \ell_L)
  = 
    \sum_{i = p}^{0} \frac{1}{\lambda^{|i|}}
          \sum_{j} G_{i j}^{[1]} \left(\ell_1, \ldots, \ell_l \right)
                G_{i j}^{[2]}\left(\ell_{l+1}, \ldots, \ell_L\right) + \mathcal{O}(\lambda).
    \label{eq:UltravioletTaylorExpansion}
\end{equation}
Here, the $i = p$ term in the summation captures the leading ultraviolet
divergence and we explicitly expand all the way to $i = 0$ which captures the
logarithmic divergence.
For each power of $\lambda$, the coefficient is a sum of terms that are
explicitly products of Feynman integrands where one factor depends only on
$\ell_1, \ldots, \ell_l$ and the other factor depends only on $\ell_{l+1},
\ldots, \ell_L$.
To see how the product structure arises, first note that it is trivially true
for all terms in the numerator as a polynomial in two sets of variables can be
written as a sum of products of polynomials in each variable. Next, consider
that the Taylor expansion of a Feynman propagator that depends on both an
ultraviolet momentum and a momentum which does not scale is an infinite sum of
terms whose denominators only depend on the ultraviolet momentum. Together, this
leads us to the decomposition of \cref{eq:UltravioletTaylorExpansion}.
This structural observation allows us to define counterterm operators for
ultraviolet singular regions. We will consider two cases, first where loop
momenta are either ultraviolet or generic, and second where one loop momentum is
ultraviolet and the other infrared.

\paragraph{Pure Ultraviolet Counterterms}
We first consider purely ultraviolet regions. If we consider truncating the
expansion in \cref{eq:UltravioletTaylorExpansion} at $\mathcal{O}(\lambda)$, we
capture the complete behavior of the integrand in the $\Gamma^{(l)}$-ultraviolet
region and have a natural candidate for a counterterm.
However, the $G_{i j}^{[1]}$ are homogeneous functions of the loop momenta
$\ell_1, \ldots, \ell_l$ and hence their integrals are scaleless.
At the logarithmic level, this is due to a cancellation of ultraviolet and
infrared divergences. We therefore see that naive use of the truncated expansion
as an ultraviolet counterterm introduces spurious infrared singularities.
To avoid the introduction of spurious infrared singularities, we define an ultraviolet
counterterm operator as
\begin{equation}
  T_{\Gamma^{(l)}} \left( I[G] \right) = I\left[ \sum_{j} \overline{G}^{[1]}_{0j} G^{[2]}_{0j}
    + \sum_{i=p}^{-1} \sum_{j} G^{[1]}_{ij} G^{[2]}_{ij} \right],
    \label{eq:UltravioletCounterterm}
\end{equation}
where
\begin{equation}
\overline{G}^{[1]}_{0j} = G^{[1]}_{0j} |_{\ell^2 \rightarrow (\ell^2 - M_{\text{UV}}^2)}.
\end{equation}
That is, to construct $\overline{G}^{[1]}_{0j}$, we take the expression for
$G^{[1]}_{0j}$ and explicitly add a mass term, $M_{\text{UV}}^2$, to each and
every inverse propagator.
Here, $M_{\text{UV}}$ acts as a separation scale between the infrared and the
ultraviolet and thus $\overline{G}^{[1]}_{0j}$ is a combination of $l$-loop
tadpole integrands.
Importantly, all denominator factors in $\overline{G}^{[1]}_{0j}$ are now
explicitly mass-regulated in the infrared.
Therefore the counterterm in \cref{eq:UltravioletCounterterm} explicitly matches
the integrand to the logarithmic level in the $\Gamma^{(l)}$ region, while not
introducing any further singularities.

Let us make a few observations about our counterterm construction. First, similar to
the construction of the collinear counterterm, we have introduced a mass scale
$M_{\text{UV}}$. Analogously to the collinear case, we stress that when applying
the counterterm formalism to a pseudo-evanescent integral, the results must
match up to $\mathcal{O}(\epsilon)$ corrections. While individual counterterms
may then depend on $M_{\text{UV}}$, the final dependence can only arise in
$\mathcal{O}(\epsilon)$ corrections. Nevertheless, the cancellation may
prove intricate.
Second, we note that, in \cref{eq:UltravioletCounterterm}, the loop momentum
integrals over $\ell_1, \ldots, \ell_l$ momenta completely factor from those
over $\ell_{l+1}, \ldots, \ell_L$. That is, the counterterms are given by
products of a massive tadpole integrals with lower-loop Feynman integrals.
Naturally, this leads to a simpler computation than the original integral.

\paragraph{Mixed Ultraviolet-Infrared Counterterms}

The second class of region for which we define an approximation operator is the
mixed ultraviolet/infrared.
The important fact that we will exploit is that, for the pseudo-evanescent
Feynman integrals under consideration, the infrared piece of the
ultraviolet/infrared region will only contribute logarithmic scaling. That is,
the power counting in the mixed region can only be power divergent, if the power
counting in the single ultraviolet region is power divergent.
With this in mind, let us reconsider the expansion in
\cref{eq:UltravioletTaylorExpansion}. We will focus our discussion on two-loop integrals.
Importantly, $G^{[2]}_{ij}(\ell_2)$ is at worst logarithmically divergent in
infrared regions.
However, for $i<0$, $G^{[2]}_{ij}(\ell_2)$ multiplies a higher power of
$\lambda$. Therefore, we must ensure that our approximation captures potential
$\lambda/\lambda$ terms.
To this end, we define our mixed ultraviolet-infrared approximation operator
associated to an infrared region $a$ as
\begin{equation}
  T_{\Gamma^{(l)}}^{a} = \sum_{i=p}^{-1} \sum_j G^{[1]}_{ij}(\ell_1) G^{[2]}_{ij}(\ell_2)
  + \sum_j \overline{G}^{[1]}_{0j}(\ell_1)  \, \hat{O}_a \! \left[  G^{[2]}_{0j}(\ell_2) \right],
  \label{eq:UVIRApproximationDefinition}
\end{equation}
where $\hat{O}_a$ is the approximation operator associated to the infrared
region $a$. Here, in order to keep track of subleading corrections in the
infrared limit associated to $\ell_2$, for $i<0$ we use the full form of
$G^{[2]}_{ij}(\ell_2)$, without approximation.
Nevertheless, as its
multiplies a scaleless integral, such terms necessarily integrate to zero.

While we see that the counterterm definition of
\cref{eq:UVIRApproximationDefinition} only requires the computation of products
of one loop integrals, it also satisfies an important property that will
simplify its application. Specifically, the operator acts on a Feynman integral
$I$ as
\begin{equation}
  T_{\Gamma^{(l)}}^{a} \! \left[ I \right] = T_{\Gamma^{(l)}} \hat{O}_a \! \left[  I\right].
  \label{eq:UVIRApproximationFactorization}
\end{equation}
This follows from the observation that the logarithmic term in
\cref{eq:UVIRApproximationDefinition} is the product of the logarithmic terms of
the ultraviolet and infrared operators. Thus, while the integrands on each side
of \cref{eq:UVIRApproximationFactorization} do not match, they differ by
scaleless integrals. An important consequence is that
\begin{equation}
(1 - T_{\Gamma^{(l)}}^{a}) (1 - T_{\Gamma^{(l)}}) (1 - \hat{O}_a) \left[ I \right] = (1 - T_{\Gamma^{(l)}}) (1 - \hat{O}_a) \left[  I\right],
\label{eq:MixedSubtractionTriviality}
\end{equation}
which follows from application of \cref{eq:UVIRApproximationFactorization} and
that the single ultraviolet and single infrared approximations are idempotent,
i.e. $T_{\Gamma^{(l)}}^2 = T_{\Gamma^{(l)}}$ and $\hat{O}_a^2 = \hat{O}_a$.
We interpret \cref{eq:MixedSubtractionTriviality} as the statement that mixed
ultraviolet-infrared subtraction is unnecessary once a Feynman integral has been
individually ultraviolet and infrared subtracted.

\subsection{Counterterm Decomposition of Pseudo-Evanescent Integrals}

Having discussed how to construct counterterms for the individual singular
regions relevant for the class of pseudo-evanescent integrals under study in
this paper, it remains to consider how these counterterms interact with each
other when multiple such singularities are present in a given integral. Again,
we restrict to the case where the subtraction of double-infrared singularities
is unnecessary.

In order to make use of our approximation operators, we consider implementing
\cref{eq:LocalCountertermConvergence} as
\begin{equation}
  \left[ \prod_{i \in \mathcal{R}}(1 - \hat{O}_i) \right] I[G^{\text{pe}}] = \mathcal{O}(\epsilon).
  \label{eq:generalPESubtraction}
\end{equation}
Here, the product is over the set of all singular regions $\mathcal{R}$ and
$\hat{O}_i$ is the approximation operator associated to a region $i$.
If we expand the product in \cref{eq:generalPESubtraction} and move all terms
involving the $\hat{O}_i$ to the right-hand side, we find an expression for the
pseudo-evanescent integral in terms of region contributions.
Nevertheless, due to the potentially large number of regions, expanding the
product can lead to a large collection of iterated approximations.
In this section, we will discuss how this can be made to simplify in
applications to two-loop pseudo-evanescent integrals under consideration.

Our first observation is that we do not need to explicitly subtract double-infrared or ultraviolet-infrared regions.
For the case of double-infrared regions, this follows by assumption, as we focus
on integrals which are power-counting convergent in double-infrared regions.
Therefore, no double-infrared approximation operators are required in
\cref{eq:generalPESubtraction}.
For the case of combined ultraviolet-infrared regions, this follows as every
mixed ultraviolet-infrared subtraction in \cref{eq:generalPESubtraction} has a
counterpart single ultraviolet and single infrared subtraction. Therefore, by
\cref{eq:MixedSubtractionTriviality}, the action of the explicit mixed term is
trivial.
We thus conclude that we are able to organize \cref{eq:generalPESubtraction}
into a product of single infrared and purely ultraviolet subtractions. That is,
we see that
\begin{equation}
    \left[ \prod_{i \in \mathcal{R}}(1 - \hat{O}_i) \right] I[G^{\text{pe}}] = (1 - \gamma_{\text{IR}})(1 - \gamma_{\text{UV}}) I[G^{\text{pe}}],
    \label{eq:simplifiedPESubtraction}
\end{equation}
where we define the infrared and ultraviolet approximation operators through
\begin{align}
  (1 - \gamma_{\text{IR}}) &= \prod_{i \in \mathcal{R}_{\text{IR}}^{(1)}}(1 - \hat{O}_i),
  \label{eq:GeneralIRSubOperator}
  \\
  (1 - \gamma_{\text{UV}}) &= \prod_{i \in \mathcal{R}_{\text{UV}}}(1 - \hat{O}_i).
  \label{eq:GeneralUVSubOperator}
\end{align}
Here, $\mathcal{R}_{\text{IR}}^{(1)}$ denotes the set of single-infrared
regions, while $\mathcal{R}_{\text{UV}}$ denotes the set of both single- and
double-ultraviolet regions.
The operator $(1-\gamma_{\text{UV}})$ can be understood as the $R$ operator of
the BPHZ formalism~\cite{Bogoliubov:1957gp,Hepp:1966eg,Zimmermann:1969jj}.
(See ref.~\cite{Herzog:2017bjx} for a recent review.)
Nevertheless, we will not need the full machinery of the $R$ operator.
Specifically, we employ that, at two loops, the ultraviolet approximation
operator can be explicitly broken down into global and sub-divergence
contributions as
\begin{equation}
    (1 - \gamma_{\text{UV}})I[G^{\text{pe}}] = (1-T_\Gamma) \prod_{\Gamma^{(1)} \subset \Gamma}(1-T_{\Gamma^{(1)}})I[G^{\text{pe}}],
    \label{eq:UVSubtractionDefinition}
\end{equation}
where we assume that $G^{\text{pe}}$ is the integrand of a two-loop Feynman
integral with graph $\Gamma$.
Here, the first term in the product subtracts any global ultraviolet divergence,
while the terms in the product over $\Gamma^{(1)}$, the ultraviolet subgraphs of
$\Gamma$, subtract the individual subdivergences.
We note that, by construction, the ultraviolet-subtracted pseudo-evanescent
integral is now power-counting finite in all ultraviolet regions, and the
behavior in any infrared region has not been worsened.
Moreover, all of the subtraction terms themselves are still pseudo-evanescent
integrals, as the approximations have inherited the $\mu_{ij}$ numerator terms
of the input integral.

While the ultraviolet operator in \cref{eq:simplifiedPESubtraction} exhibits a
limited number of terms, the infrared operator $\gamma_{\text{IR}}$ naively
contains a great many terms.
Let us consider how it simplifies when applied to
to the class of two-loop pseudo-evanescent Feynman integrals that we consider
in this paper. Importantly, as the result of ultraviolet subtraction is a
pseudo-evanescent integral with the same infrared powercounting as the input
we can apply the discussion of \cref{sec:EvanescentPowerCounting}, assuming all
singularities are logarithmic.
As the set of regions in \cref{eq:GeneralIRSubOperator} only involves single
infrared regions, we can write the subtraction as
\begin{equation}
  (1 - \gamma_{\text{IR}})I[G^{\text{pe}}] = \prod_{k} (1 - \mathcal{C}_k)\prod_{i,j}(1 - \mathcal{S}_{(i,j)})I[G^{\text{pe}}].
  \label{eq:DoubleIRSafeIROperator}
\end{equation}
That is, we must only perform infrared subtractions in single soft and collinear
regions.
In \cref{eq:DoubleIRSafeIROperator}, the product over $k$ is taken over all
massless external legs, while the product over $i,j$ is taken over all pairs of
distinct on-shell external legs.

Our first observation is that the products of soft operators on the right-hand
side of \cref{eq:DoubleIRSafeIROperator} automatically cancel in the case of
two-loop pseudo-evanescent integrals.
To see this, without loss of generality, we assume that $\mathcal{S}_{(i,j)}$
approximates a soft divergence associated to the loop momentum $\ell_1$. As the
triangle integral factor in \cref{eq:LogarithmicSoftCounterterm} only exhibits
logarithmic power-counting in the $i,j$ region, soft subtraction in some other
region must can only give non-zero if it acts on the sub-loop associated to
$\ell_2$.
However, for the result of $\mathcal{S}_{(i,j)}I[G^{\text{pe}}]$ to
have been non-zero, it must be the case that the numerator of $G^{\text{pe}}$
contained a factor of $\mu_{22}$. This factor provides a sufficient suppression
in any soft region on the $\ell_2$ loop, such that the action of a further soft
subtraction must be zero. We therefore conclude that the iterated soft
subtraction does not generate a proliferation of terms and we can write that
\begin{equation}
  \label{eq:SoftProliferationTreaty}
  \prod_{i,j}\left( 1 - \mathcal{S}_{(i,j)} \right) I[G^{\text{pe}}] = \left( 1 - \mathcal{S} \right) I[G^{\text{pe}}],
\end{equation}
where
\begin{equation}
\mathcal{S} = \sum_{i, j} \mathcal{S}_{(i,j)}.
  \label{eq:FullSoftDefinition}
\end{equation}
It is not hard to see that an analogous argumentation applied to the case of
collinear counterterms leads to a similar lack of iterated collinear
subtractions. That is, we have that
\begin{equation}
  \left[\prod_k \left( 1 - \mathcal{C}_k \right) \right] I[G^{\text{pe}}] = \left( 1 - \mathcal{C} \right) I[G^{\text{pe}}],
\end{equation}
where we define
\begin{equation}
\mathcal{C} = \sum_k \mathcal{C}_k.
\end{equation}
If we consider the combined infrared subtraction, we see that we can write it as
\begin{equation}
    \gamma_{\text{IR}} I[G^{\text{pe}}] = \mathcal{S} + \mathcal{C}(1-\mathcal{S})I[G^{\text{pe}}].
\end{equation}
Importantly, the collinear approximation acts non-trivially on the soft
approximation.  This follows as the triangle integral in
\cref{eq:LogarithmicSoftCounterterm} may itself have divergent power counting in
the collinear region.
Nevertheless, it should be clear that this interaction is mild. Each soft
subtraction term has collinear divergences associated to the legs $i$ and $j$.
Therefore, the collinear-subtracted soft-subtraction term is only non-zero if
either $i$ or $j$ is $k$. That is,
\begin{equation}
  \mathcal{C}_k \mathcal{S}_{(i,j)} = \mathcal{C}_k \mathcal{S}_{(i,j)} (\delta_{i,k} + \delta_{j,k}),
\end{equation}
where only one term contributes on the right-hand side as $i \ne j$. This leads us to define
\begin{equation}
    \mathcal{C}^{(+)} = \sum_k \mathcal{C}^{(+)}_{k}, \qquad \qquad \mathcal{C}^{(+)}_{k} = \mathcal{C}_{k} \sum_l (1- \mathcal{S}_{(k,l)}).
\end{equation}
Here we embellish the collinear operator with a plus, as we will later see that the 
the overlap contribution gives rise to a plus prescription. In analogy with \cref{eq:collienearSsub}, we define $\overline{\mathcal{C}}^{(+)}_k$ by removing the same factors.
Altogether, we see that we can simplify the action of $\gamma_{\text{IR}}$ on a
pseudo-evanescent integral to the form
\begin{equation}
  \gamma_{\mathrm{IR}}I[G^{\text{pe}}] = \left( \mathcal{S} + \mathcal{C}^{(+)} \right)I [G^{\text{pe}}].
  \label{eq:IRMuCountertermDefinition}
\end{equation}
That is, the infrared contribution is given by a sum of soft contributions and (plus-prescribed) collinear contributions.

In summary, this discussion tells us that we can implement
\cref{eq:MuCountertermEquivalence} by directly expressing a pseudo-evanescent
integral in terms of the local counterterms as
\begin{equation}
  I[G^{\text{pe}}] = \left(\gamma_{\mathrm{IR}} + [1 - \gamma_{\mathrm{IR}}] \gamma_{\mathrm{UV}}\right) I [G^{\text{pe}}] + \mathcal{O}(\epsilon),
  \label{eq:FullMuCountertermDefinition}
\end{equation}
where the action of $\gamma_{\mathrm{UV}}$ is defined in \cref{eq:UVSubtractionDefinition} and
the action of $\gamma_{\mathrm{IR}}$ is defined in \cref{eq:IRMuCountertermDefinition}. 
These two operators give ultraviolet and infrared contributions respectively,
and in \cref{eq:FullMuCountertermDefinition} there is further a
contribution related to their overlap.
Similarly in \cref{eq:IRMuCountertermDefinition}, the overlap of the
soft and collinear regions must also be appropriately subtracted.

\paragraph{Example Decomposition}
Let us now present an example to showcase our decomposition of pseudo-evanescent integrals. We will use a two-loop massless double box integral with a $\mu_{22}^2$ insertion. Specifically, we use the integral whose region structure we described in \cref{fig:regionsdivdb}. Notably, (up to permutations) this integral is the only non-factorizable integral contributing to the leading-color two-loop $4$-point all-plus amplitude~\cite{Bern:2000dn}. As such, the following discussion can easily be promoted to a full computation of the leading-color amplitude, which we leave as an exercise for the reader.
Let us recall the possible singular regions that have associated subtraction terms from \cref{fig:regionsdivdb}. We recall that we only have infrared-singular power counting in regions associated to the loop momentum $\ell_1$: a soft singularity associated to the exchange between $p_1$ and $p_2$, as well as collinear singularities associated to $p_1$ and $p_2$. 
On the side of ultraviolet-singular regions, the integral is singular in the large $\ell_2$ limit. By \cref{eq:FullMuCountertermDefinition}, we therefore can write
\begin{equation}
   \eqnDiagr[8pt]{\includegraphics[scale=0.5]{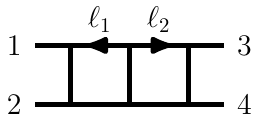}}[\mu_{22}^2] = \left(\mathcal{S}_{(1,2)}+\mathcal{C}^{(+)}_{1}+\mathcal{C}^{(+)}_{2} +  [1 - \gamma_{\text{IR}}]T_{\Gamma_2^{(1)}} \right) \eqnDiagr[8pt]{\includegraphics[scale=0.5]{graphics/doubleBOX.pdf}}[\mu_{22}^2] + \mathcal{O}(\epsilon),
   \label{eq:DBRegionDecomposition}
\end{equation}
where $\Gamma_2^{(1)}$ is the one-loop sub-diagram associated to the $\ell_2$ loop.
That is, we receive a contribution from each singular infrared region as well as a contribution from the ultraviolet region, which must then be infrared subtracted.

Let us consider each of these contributions in turn, starting by analyzing the soft contribution. Considering the counterterm definition of \cref{eq:LogarithmicSoftCounterterm}, the contribution is given by the product of a one-loop triangle and the remaining sub-diagram evaluated with $\ell_1 = p_1$. This sub-diagram is easily identifiable as a one-loop box diagram and as such we can write the contribution as 
\begin{equation}
    \mathcal{S}_{(1,2)}\left( \eqnDiagr[8pt]{\includegraphics[scale=0.5]{graphics/doubleBOX.pdf}}[\mu_{22}^2] \right)= \eqnDiagr[4pt]{\includegraphics[scale=0.50]{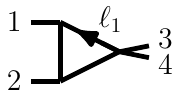}} \eqnDiagr[10pt]{\includegraphics[scale=0.5]{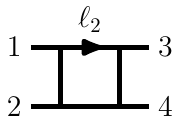}}[\mu_{22}^2].
\end{equation}
Importantly, the soft contribution is given by a product of one-loop integrals and is therefore significantly easier to compute than the full two-loop double box.
Let us now consider the computation of the collinear contributions. Clearly the diagram is symmetric and we must only consider a single contribution. First, note that the $\mathcal{C}^{(+)}_1$ is defined to collinear approximate the soft-subtracted double box. 
As we have already computed the soft counterterm, it is easy to apply \cref{eq:collienearSsub} to find
\begin{equation}
    \overline{\mathcal{C}}_{1}^{(+)}\left( \eqnDiagr[8pt]{\includegraphics[scale=0.5]{graphics/doubleBOX.pdf}}[\mu_{22}^2] \right) 
    = 
    \frac{1}{s_{12} \epsilon} \int_0^1 \mathrm{d} x \left[\frac{1}{1-x}\right] \left(
    \eqnDiagr[4pt]{\includegraphics[scale=0.5]{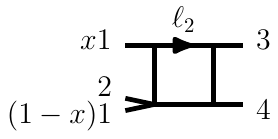}}[\mu_{22}^2]  - \eqnDiagr[8pt]{\includegraphics[scale=0.5]{graphics/box4pl2.pdf}}[\mu_{22}^2] \right) \,.
\end{equation}
Here, the first term is associated to the collinear approximation of the double box, while the second is associated to the collinear approximation of the soft subtraction term. In both terms, the remaining sub-diagram has been evaluated on the collinear configuration $\ell_1 = x p_1$. The factor of $\frac{1}{s_{12}(1-x)}$ arises in both terms from evaluating the hard propagator in the $\ell_1$ sub-loop on the collinear configuration.
Interestingly, the massless box can be seen to be the $x = 0$ limit of the $x$-dependent box. As such, we can express the collinear counterterm simply as 
\begin{equation}
    \overline{\mathcal{C}}_{1}^{(+)}\left( \eqnDiagr[8pt]{\includegraphics[scale=0.5]{graphics/doubleBOX.pdf}}[\mu_{22}^2] \right)= \frac{1}{s_{12} \epsilon} \int_0^1 \mathrm{d} x \left[\frac{1}{1-x}\right]_+ \eqnDiagr[4pt]{\includegraphics[scale=0.5]{graphics/boxl2C1.pdf}}[\mu_{22}^2]  \,,
\end{equation}
where we have made use of the plus distribution to subtract the end-point contribution at $x = 1$.
We note that the occurence of a plus distribution to regulate end-point singularities is a common feature of $\overline{\mathcal{C}}_i^{(+)}$, which motivates the superscript.
Finally we can study the ultraviolet contribution coming from the large $\ell_2$ region. This can be constructed following \cref{eq:UltravioletCounterterm} as
\begin{equation}
    T_{\Gamma^{(1)}_2}\left( \eqnDiagr[8pt]{\includegraphics[scale=0.5]{graphics/doubleBOX.pdf}}[\mu_{22}^2] \right)=  \eqnDiagr[2pt]{\includegraphics[scale=0.50]{graphics/tril1.pdf}}  \eqnDiagr{\includegraphics[scale=0.30]{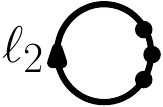}}[\mu_{22}^2] \,,
\end{equation}
which takes a simple form as the ultraviolet singularity is only logarithmic so the Taylor expansion around the limit can be truncated at first order.
Following the prescription of \cref{eq:FullMuCountertermDefinition} we need to subtract the overlap between the soft and ultraviolet regions which amounts to
\begin{equation}
\mathcal{S}_{(1,2)}\left( \eqnDiagr[2pt]{\includegraphics[scale=0.50]{graphics/tril1.pdf}}  \eqnDiagr{\includegraphics[scale=0.30]{graphics/quadTadpole2.pdf}}[\mu_{22}^2] \right)= \eqnDiagr[2pt]{\includegraphics[scale=0.50]{graphics/tril1.pdf}}  \eqnDiagr{\includegraphics[scale=0.30]{graphics/quadTadpole2.pdf}}[\mu_{22}^2] \,.
\end{equation}
That is, the soft counterterm of the ultraviolet contribution is exactly the ultraviolet contribution. Therefore, the soft-subtracted ultraviolet counterterm is actually zero in this case. Naturally, performing collinear subtraction on this result gives zero and we conclude that the infrared subtracted ultraviolet contribution vanishes, i.e.
\begin{equation}
    (1-\gamma_{\text{IR}}) T_{\Gamma^{(1)}_2}\left( \eqnDiagr[8pt]{\includegraphics[scale=0.5]{graphics/doubleBOX.pdf}}[\mu_{22}^2] \right) = 0,
\end{equation}
providing the final contribution to \cref{eq:DBRegionDecomposition}.

Combining all of the contributions to \cref{eq:DBRegionDecomposition}, explicitly we obtain the physically transparent result
\begin{align}
\begin{split}
\eqnDiagr[8pt]{\includegraphics[scale=0.5]{graphics/doubleBOX.pdf}}[\mu_{22}^2]
=&\eqnDiagr[2pt]{\includegraphics[scale=0.50]{graphics/tril1.pdf}} \eqnDiagr[8pt]{\includegraphics[scale=0.50]{graphics/box4pl2.pdf}}[\mu_{22}^2] 
 \\ &+ \frac{1}{s_{12} \epsilon} \int_0^1 \mathrm{d} x \left[\frac{1}{1-x}\right]_+ \eqnDiagr[4pt]{\includegraphics[scale=0.5]{graphics/boxl2C1.pdf}}[\mu_{22}^2]+\left(\genfrac{}{}{0pt}{}{1 \rightarrow 2}{3 \rightarrow 4}\right) + \mathcal{O}(\epsilon) \,.
\end{split}
\label{eq:doubleboxTot}
\end{align} 
This decomposition is easily tested by explicitly computing the integrals appearing in \cref{eq:doubleboxTot}. The double box integral can be reduced to master integrals using tools such as ref.~\cite{Smirnov:2025prc} and expressed in terms of well-known master integrals~\cite{Gehrmann:1999as,Anastasiou:1999cx,Anastasiou:1999bn,Smirnov:1999wz,Anastasiou:2000kp}. The integrals on the right-hand side can be computed using \texttt{HyperInt}~\cite{Panzer:2014caa} after making use of dimension-shifting relations. In this way, we have computationally confirmed \cref{eq:doubleboxTot}.
\section{The Two-Loop 5-Point All-Plus Amplitude from Counterterms}

As an application of our methodology, we will now consider the full-color,
five-point two-loop all-plus amplitude. The complete integrands for this
amplitude were first presented in \cite{Badger:2015lda}, and were integrated by
traditional means in \cite{Badger:2019djh}. Moreover, an approach based on
one-loop unitarity and augmented recursion was used to compute the amplitudes in
\cite{Dunbar:2019fcq}. In our approach, we exploit that the two-loop all-plus
amplitudes are given by a linear combination of pseudo-evanescent integrals.
This provides a non-trivial, physical example to stress test our approach.

\subsection{Notation}

We denote the perturbative expansion in terms of the bare coupling $\alpha_0$ as
\begin{equation}
  \mathcal{A}_5 = {\left( 4\pi\alpha_0 \right)}^{\frac{3}{2}}
  \sum_{l=0}^\infty {\left( \frac{\alpha_0}{2\pi} \right)}^{l} \mathcal{A}_5^{(l)}.
\end{equation}
As usual, we denote the external momenta as $p_i$, taking external momenta to be
outgoing. A sum of external momenta is represented as $p_{a \ldots b} = p_a +
\cdots + p_b$.
The one-loop amplitude is well known~\cite{Bern:1993mq}, and we make use of the
form given in~\cite{Badger:2016ozq}.
\begin{equation}
    \mathcal{A}_5^{(1)} = -(D_s- 2) \sum_{\sigma \in S_5/D_5} \sigma \circ \Bigg[ C\left(\!\! \eqnDiag{\includegraphics[scale=0.25]{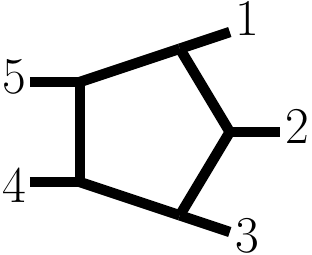}} \! \right)
    \frac{\overline{A}^{(1)}_5}{\langle 12 \rangle \langle 23 \rangle \langle 34 \rangle \langle 45 \rangle \langle 51 \rangle} 
  \Bigg] .
  \label{eq:OneLoopColourDecomposition}
\end{equation}
where
\begin{equation}
    \overline{A}^{(1)}_5 \!=
    \!\! \eqnDiag{\includegraphics[scale=0.28]{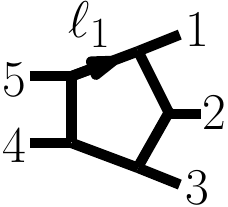}} \! \left[\mu_{11}^2 \mathrm{tr}_+(1(\ell_1\!-\!p_1)(\ell_1\!-\!p_{12})345)\right]
    - \eqnDiag{\includegraphics[scale=0.28]{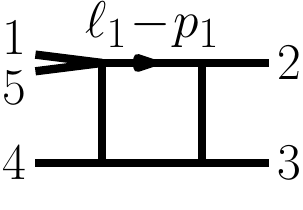}} [\mu_{11}^2 s_{23}s_{34}]
      - \eqnDiag{\includegraphics[scale=0.38]{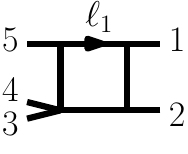}} [\mu_{11}^2 s_{12}s_{15}],
\end{equation}
and the sum is over all the 12 non-cyclic permutations that are not related by
reflection. Specifically, the permutation $\sigma$ acts on all external leg labels: momentum, spinor and color.
The color factor in \cref{eq:OneLoopColourDecomposition} is denoted
diagrammatically as $C$ acting on a diagram. The diagram specifies a
color-factor expression in terms of color Feynman rules.
As we consider a purely gluonic amplitude, the color factors involved are all
composed of adjoint generators and each vertex in the diagram represents a
factor of some $f_{abc}$.
As an explicit example, if to the $i^{\text{th}}$ external gluon we associate
the adjoint index $a_i$, the color factor in
\cref{eq:OneLoopColourDecomposition} explicitly reads
\begin{equation}
  C\left(\!\! \eqnDiag{\includegraphics[scale=0.25]{graphics/pentagonColor.pdf}} \! \right) = \tilde{f}^{e_{1} a_{1} e_{2}} \tilde{f}^{e_{2} a_{2} e_{3}} \tilde{f}^{e_{3} a_{3} e_{4}} \tilde{f}^{e_{4} a_{4} e_{5}} \tilde{f}^{e_{5} a_{5} e_{1}},
\end{equation}
where the adjoint color factors $\tilde{f}^{abc}$ are defined through
\begin{equation}
  \tilde{f}^{abc} = \sqrt{2} i f^{abc} = \mathrm{Tr}([T^a, T^b] T^c),
  \qquad \qquad 
  \mathrm{Tr}(T^a T^b) = \delta^{ab},
\end{equation}
where $T^a$ are the fundamental generators of $\text{SU}(N)$.
We refer the reader to~\cite{Ochirov:2016ewn,Ochirov:2019mtf} for further details
on the diagrammatic notation.

Our main point of focus is the five-point two-loop all-plus amplitude. We make
use of the integrand of \cite{Badger:2015lda}, from which we take the
conventions. The amplitude can be written as
\begin{align}
  \begin{split}
     \mathcal{A}^{(2)}_5 = \,\,\,\, 
    \sum_{\sigma \in S_5} \sigma \!\circ\!\Bigg[ 
    &C\!\left(\!\!\! \eqnDiag{\includegraphics[scale=0.30]{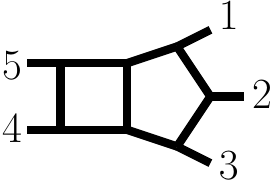}} \! \right)\!
    \Bigg\{ \!\! \eqnDiag{\includegraphics[scale=0.30]{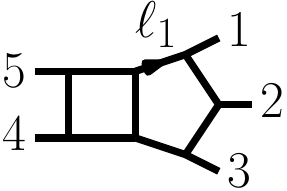}}\left[\frac{F_1 N_{\text{pb}}(\ell_1, \ell_2)}{2} \right]
      + \eqnDiag{\includegraphics[scale=0.30]{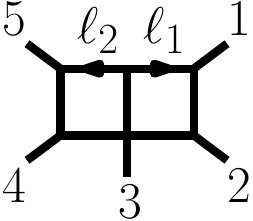}}\left[ \frac{F_1 N_{\text{ssdb}}}{2} \right]
    \\
      & \hspace{28mm} + \eqnDiag{\includegraphics[scale=0.30]{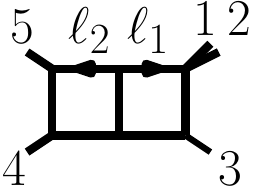}}[ F_1 N_{\text{1mdb}}] +  
     \eqnDiag{\includegraphics[scale=0.25]{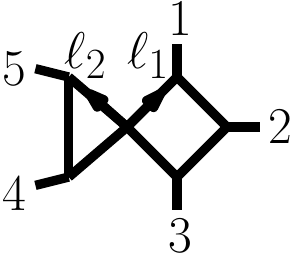}}\left[\frac{N_{\text{bt}}(\ell_1, \ell_2)}{2}\right]
     \\
      & \hspace{28mm} + \eqnDiag{\includegraphics[scale=0.25]{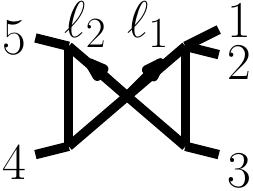}}[N_{\text{tt1m}}(\ell_1, \ell_2)]
      + \eqnDiag{\includegraphics[scale=0.25]{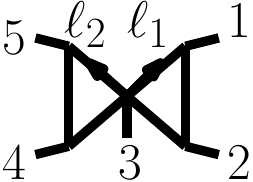}}\left[\frac{N_{\text{sstt}}(\ell_1, \ell_2)}{2} \right] \Bigg\} + \\
      &C\left(\!\!
        \eqnDiag{\includegraphics[scale=0.45]{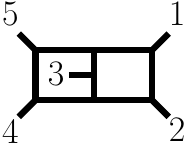}}\right)
        \Bigg(
        \! \eqnDiag{\includegraphics[scale=0.45]{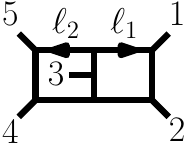}}\left[ \frac{F_1 N_{\text{dp}}(\ell_1, \ell_2)}{4} \right]
        + \eqnDiag{\includegraphics[scale=0.45]{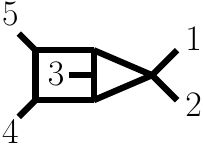}}\left[ \frac{F_1 N_{\text{nppb}}}{2} \right]
        \\
        &\hspace{28mm} + \eqnDiag{\includegraphics[scale=0.35]{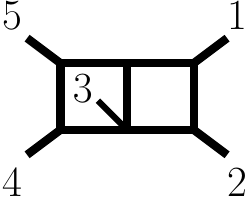}}\left[ \frac{F_1 N_{\text{npdb}}}{2} \right] 
          - \eqnDiag{\includegraphics[scale=0.25]{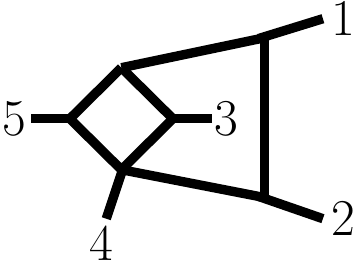}}\left[ F_1 N_{\text{pbx}} \right]
        \\
        & \hspace{28mm} + \eqnDiag{\includegraphics[scale=0.25]{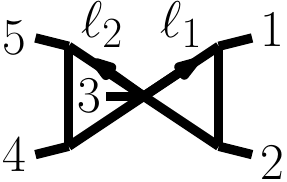}}\left[ \frac{F_1 N_{\text{npsstt}}(\ell_1, \ell_2)}{4} \right]
      \Bigg) + \\
      &C\left( \!\!\! \eqnDiag{\includegraphics[scale=0.25]{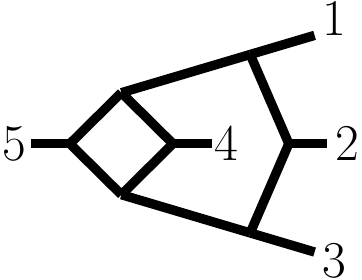}}\right) \left( \eqnDiag{\includegraphics[scale=0.25]{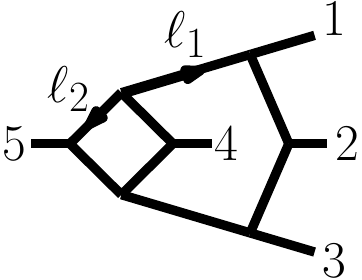}} \left[ \frac{F_1 N_{\text{hb}}(\ell_1)}{4} \right] + \eqnDiag{\includegraphics[scale=0.25]{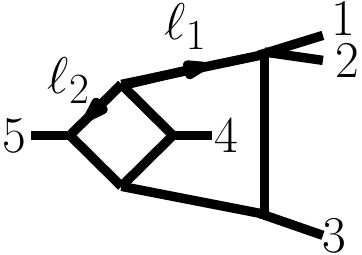}} \left[ \frac{F_1 N_{\text{1mpx}}}{2} \right] \right)
    \Bigg],
  \end{split}
  \label{eq:AllPlusFivePointAmplitude}
\end{align}
where we abbreviate the frequently arising pseudo-evanescent pre-factor
\begin{equation}
  F_1 = (D_s-2)(\mu_{11}\mu_{22} + (\mu_{11}+ \mu_{22})^2 + 2 \mu_{12}(\mu_{11} + \mu_{22})) + 16 (\mu_{12}^2 - \mu_{11}\mu_{22}).
\end{equation}
Explicit expressions for the numerators can be found in 
\cref{sec:AllPlusFivePointIntegrand}. In contrast to ref.~\cite{Badger:2015lda}, we
note that we have defined a number of our numerators with the factor of
$F_1$ explicitly pulled out. As we will see, this factor plays a special role in
our formalism.

\subsection{Amplitude Computation}
In the following, we apply the counterterm formalism to the complete all-plus
amplitude. Importantly, as the amplitude is a linear combination of
(pseudo-)evanescent integrals, we can apply
\cref{eq:FullMuCountertermDefinition}, taking $G^{\text{pe}}$ to be the
integrand of \cref{eq:AllPlusFivePointAmplitude}. This allows us to decompose
the amplitude into a sum of region contributions. Specifically, we decompose as
\begin{equation}
  \mathcal{A}_5^{(2)} =
  \mathcal{A}_5^{(2),\text{soft}} + \mathcal{A}_5^{(2),\text{col}}+ \mathcal{A}_5^{(2),\text{UV}} + \mathcal{O}(\epsilon),
\end{equation}
where we define
\begin{equation}
  \mathcal{A}_5^{(2),\text{soft}} = \mathcal{S}[\mathcal{A}_5^{(2)}], \quad\,\,
  \mathcal{A}_5^{(2),\text{col}} = \mathcal{C}^{(+)}[\mathcal{A}_5^{(2)}] \,\,\quad \text{and} \quad\,\,
  \mathcal{A}_5^{(2),\text{UV}} = (1 \!-\! \gamma_{\mathrm{IR}}) \gamma_{\mathrm{UV}} [\mathcal{A}_5^{(2)}].
\end{equation}
We refer to the first contribution as the soft contribution, the second as the
collinear contribution, and the third contribution as the (infrared-subtracted)
ultraviolet contribution.

\subsubsection{Soft Contribution}

Here we compute the soft contribution to the amplitude,
$\mathcal{A}_5^{(2),\text{soft}}$. Let us begin by noting that we can split the
soft contribution to the amplitude into a sum over contributions each
associated to a different soft exchange. That is, we write
\begin{equation}
  \mathcal{A}_5^{(2),\text{soft}} = \sum_{1 \le i < j \le n} \mathcal{A}_5^{(2),[i,j]\text{-soft}}, \qquad \qquad \mathcal{A}_5^{(2),[i,j]\text{-soft}} = \mathcal{S}_{(i,j)} \left[ \mathcal{A}_5^{(2)} \right].
  \label{eq:AmplitudeSoftDecomposition}
\end{equation}
This decomposition will help us organize the soft contribution in a physically
transparent way.
Note that the soft contribution of each summand in
\cref{eq:AllPlusFivePointAmplitude} will contribute to multiple
$\mathcal{A}_5^{(2),[i,j]\text{-soft}}$.
Only through the final permutation sum in \cref{eq:AllPlusFivePointAmplitude} do
we recover a form for $\mathcal{A}_5^{(2),[i,j]\text{-soft}}$.
We, therefore, proceed by systematically computing the soft counterterms for each
summand of \cref{eq:AllPlusFivePointAmplitude}.
As a first observation, we note that all factorizable graphs in
\cref{eq:AllPlusFivePointAmplitude} are convergent in single-soft regions.
Their soft contributions are therefore zero, and we do not discuss them further.
We, thus, begin with the non-factorizable, planar graphs. The soft contribution
to the pentabox topology reads
\begin{align}
  \begin{split}
    \mathcal{S} \! &\left( \!\!\!
      \eqnDiag{\includegraphics[scale=0.30]{graphics/scalarPentaboxBadgerLabel.pdf}}[F_1 N_{\text{pb}}(\ell_1, \ell_2)]
      \! \right) \\
    &= (D_s \!-\! 2)\Bigg\{
    \!\!\! \eqnDiag{\includegraphics[scale=0.3]{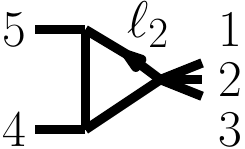}}[s_{45}] \eqnDiag{\includegraphics[scale=0.28]{graphics/pentagonScalarL1.pdf}}\left[\frac{\mu_{11}^2N_{\text{pb}}(\ell_1, p_5)}{s_{45}}\right] 
    \!+\!
    \eqnDiag{\includegraphics[scale=0.3]{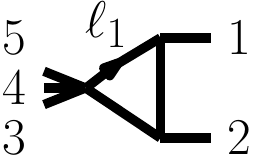}}[s_{12}] \!\eqnDiag{\includegraphics[scale=0.42]{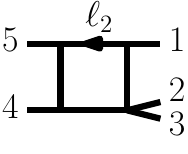}}\left[ \frac{\mu_{22}^2 N_{\text{pb}}(p_1, \ell_2)}{s_{12}s_{23}} \right]\\
    &\qquad \qquad \,\,\quad + 
    \eqnDiag{\includegraphics[scale=0.3]{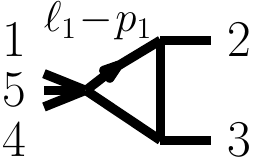}}[s_{23}]
    \eqnDiag{\includegraphics[scale=0.42]{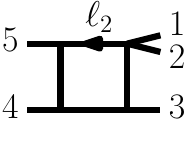}}\left[\frac{\mu_{22}^2N_{\text{pb}}(p_{12}, \ell_2)}{s_{23}s_{12}}\right] \!\! \Bigg\},
  \end{split}
  \label{eq:AllPlusPentaboxSoft}
\end{align}
where we explicitly see that all soft exchanges in the pentabox contribute.
In \cref{eq:AllPlusPentaboxSoft}, and in later soft contributions, for each
one-mass triangle we judiciously keep a factor of the associated invariant in
the numerator.
As we will see later, this choice is natural when considering the universal soft
behavior of the amplitude. The remaining soft contributions to each planar piece
of the integrand are given by
\begin{align}
  \begin{split}
    &\mathcal{S}\left(\!\!\eqnDiag{\includegraphics[scale=0.3]{graphics/ssdbBadgerLabel.pdf}}[ F_1 N_{\text{ssdb}}]\!\right) \\
    &= (D_s - 2) \Bigg\{\!\!\!
    \eqnDiag{\includegraphics[scale=0.3]{graphics/S45L2Triangle.pdf}}[s_{45}] \! \eqnDiag{\includegraphics[scale=0.40]{graphics/S34BoxL1.pdf}}\!\left[\frac{\mu_{11}^2 N_{\text{ssdb}}}{s_{45}}\right]
    +
    \eqnDiag{\includegraphics[scale=0.3]{graphics/S12L1Triangle.pdf}}[s_{12}] \! \eqnDiag{\includegraphics[scale=0.40]{graphics/S23BoxL2.pdf}}\!\left[\frac{\mu_{22}^2 N_{\text{ssdb}}}{s_{12}}\right]
    \!\!\Bigg\}.
  \end{split}
      \\
    &\mathcal{S}\left(\!\!\eqnDiag{\includegraphics[scale=0.3]{graphics/oneMassDoubleBoxBadgerLabel.pdf}}[ F_1 N_{\text{1mdb}}]\!\right) 
    = (D_s - 2) \Bigg\{\!\!\!
    \eqnDiag{\includegraphics[scale=0.3]{graphics/S45L2Triangle.pdf}}[s_{45}] \eqnDiag{\includegraphics[scale=0.40]{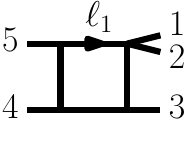}}\!\left[\frac{\mu_{11}^2 N_{\text{1mdb}}}{s_{45}}\right]
    \!\!\Bigg\}.
\end{align}
Let us note that it is possible to read from these expressions the explicit soft
exchange to which each piece belongs directly from the one-loop triangle
integral that is involved.

Let us now shift to the non-planar contributions. Looking to
\cref{eq:AllPlusFivePointAmplitude}, we see that they are grouped into two
different color factors.
For the non-planar hexagon box color factor, only the first term is soft
divergent and so the only contribution is
\begin{align}
  \begin{split}
    &\mathcal{S}\left(\!\!\eqnDiag{\includegraphics[scale=0.25]{graphics/hexabox.pdf}} \left[ \frac{F_1 N_{\text{hb}}(\ell_1)}{4} \right]\!\right) \\
    &= (D_s - 2) \Bigg\{\!\!\!
    \eqnDiag{\includegraphics[scale=0.3]{graphics/S12L1Triangle.pdf}}[s_{12}] \! \eqnDiag{\includegraphics[scale=0.25]{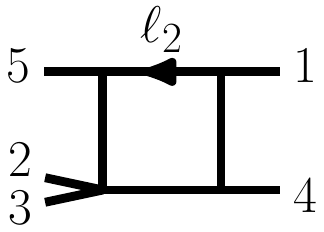}}\!\left[\frac{\mu_{22}^2 N_{\text{hb}}(p_1)}{s_{12}}\right]
    +
    \eqnDiag{\includegraphics[scale=0.3]{graphics/S23L1Triangle.pdf}}[s_{23}] \! \eqnDiag{\includegraphics[scale=0.25]{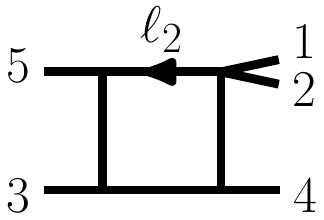}}\!\left[\frac{\mu_{22}^2 N_{\text{hb}}(p_{12})}{s_{23}}\right]
    \!\!\Bigg\}.
  \end{split}
\end{align}
The double-pentagon color factor multiplies a larger number of soft divergent
contributions. Each of the double pentagon topology and five-point double box
topology gives two contributions as
\begin{align}
  \begin{split}
    &\mathcal{S} \! \left( \!
      \eqnDiag{\includegraphics[scale=0.45]{graphics/doublePentagonBadgerLabel.pdf}}[F_1 N_{\text{dp}}(\ell_1, \ell_2)]
      \! \right) \\
    &= (D_s \!-\! 2)\Bigg\{
    \!\!\! \eqnDiag{\includegraphics[scale=0.28]{graphics/S45L2Triangle.pdf}}[s_{45}] \! \eqnDiag{\includegraphics[scale=0.21]{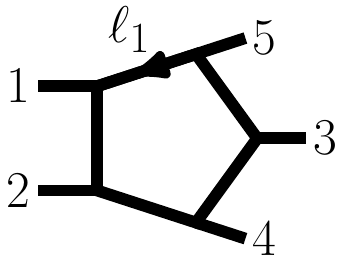}} \! \left[\frac{\mu_{11}^2N_{\text{dp}}(\ell_1, p_5)}{s_{45}}\right]
    +
    \eqnDiag{\includegraphics[scale=0.28]{graphics/S12L1Triangle.pdf}}[s_{12}] \! \eqnDiag{\includegraphics[scale=0.21]{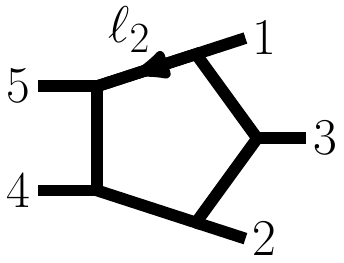}} \! \left[\frac{\mu_{22}^2N_{\text{dp}}(p_1,\ell_2)}{s_{12}}\right] \!\! \Bigg\},
  \end{split}
  \\
  \begin{split}
    &\mathcal{S}\left(\!\!\eqnDiag{\includegraphics[scale=0.35]{graphics/npdb.pdf}}[ F_1 N_{\text{npdb}}]\!\right) \\
    &= (D_s - 2) \Bigg\{\!\!\!
    \eqnDiag{\includegraphics[scale=0.3]{graphics/S45L2Triangle.pdf}}[s_{45}] \! \eqnDiag{\includegraphics[scale=0.40]{graphics/S34BoxL1.pdf}}\!\left[\frac{\mu_{11}^2 N_{\text{npdb}}}{s_{45}}\right]
    +
    \eqnDiag{\includegraphics[scale=0.3]{graphics/S12L1Triangle.pdf}}[s_{12}] \! \eqnDiag{\includegraphics[scale=0.40]{graphics/S23BoxL2.pdf}}\!\left[\frac{\mu_{22}^2 N_{\text{npdb}}}{s_{12}}\right]
    \!\!\Bigg\}.
  \end{split}
\end{align}
The remaining two non-planar topologies give rise to the final
non-zero contributions to the soft region, which are
\begin{align}
    &\mathcal{S}\left(\!\!\eqnDiag{\includegraphics[scale=0.45]{graphics/pentagonBoxNP.pdf}}[ F_1 N_{\text{nppb}}]\!\right) 
    = (D_s - 2) \Bigg\{\!\!\!
    \eqnDiag{\includegraphics[scale=0.3]{graphics/S45L2Triangle.pdf}}[s_{45}] \eqnDiag{\includegraphics[scale=0.25]{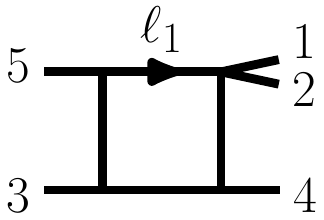}}\!\left[\frac{\mu_{11}^2 N_{\text{nppb}}}{s_{45}}\right]
    \!\!\Bigg\}. \\
    &\mathcal{S}\left(\!\!\eqnDiag{\includegraphics[scale=0.25]{graphics/pbx.pdf}}[ F_1 N_{\text{pbx}}]\!\right) 
    = (D_s - 2) \Bigg\{\!\!\!
    \eqnDiag{\includegraphics[scale=0.3]{graphics/S12L1Triangle.pdf}}[s_{12}] \eqnDiag{\includegraphics[scale=0.25]{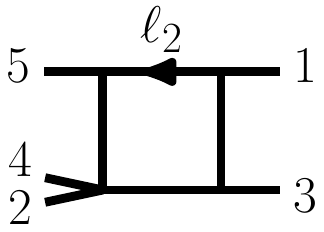}}\!\left[\frac{\mu_{22}^2 N_{\text{pbx}}}{s_{12}}\right]
    \!\!\Bigg\}.
\end{align}

With all these ingredients in hand, next, we combine the pieces and gather only
those that contribute to a single soft exchange. For concreteness, we focus
on $\mathcal{A}_5^{[1,2]-\text{soft}}$.
After some non-trivial color algebra, but only making use of integrand level
relations, we find that
\begin{equation}
  \mathcal{A}_5^{(2),[1,2]\text{-soft}} \!= \!\!\eqnDiag{\includegraphics[scale=0.30]{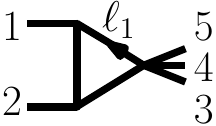}}[s_{12}] \!\!\!\!\! \sum_{\sigma \in S[3,4,5]} \!\!\!\!\!\! \sigma \circ \left[ C\!\left(\!\!\! \eqnDiag{\includegraphics[scale=0.30]{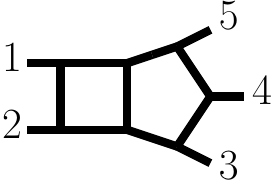}} \!\! \right) \!  A_5^{(1)}(1,\!2,\!3,\!4,\!5) + C\!\left(\!\!\eqnDiag{\includegraphics[scale=0.37]{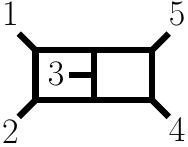}} \right) \! A_5^{(1)}(1,\!3,\!2,\!4,\!5) \! \right]\!,
  \label{eq:SingleSoftExchangeContribution}
\end{equation}
where $S[3,4,5]$ is the set of $3!$ permutations of legs $3,4,5$.
We stress that the remaining soft exchanges are fixed by the fact that the full
all-plus amplitude is completely Bose symmetric.

Interestingly, we see that an individual soft contribution can explicitly be
written in terms of the one-loop amplitude. Indeed, appropriately written, it
turns out that \cref{eq:SingleSoftExchangeContribution} expresses that a single
soft contribution to the amplitude factorizes.
To manifest this, we define a single-soft insertion operator that is adjusted to
our local subtraction. Specifically, we introduce
\begin{equation}
  \overline{\mathbfcal{Z}}_n^{(1), [i,j]\text{-soft}} = {\bf T}_i \cdot {\bf T}_j \eqnDiag{\includegraphics[scale=0.4]{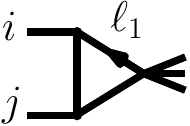}}[s_{ij}],
\end{equation}
where ${\bf T}_i$ is the adjoint color insertion operator associated to
particle $i$ and we
suppress the other momentum labels in the triangle integral.
This allows us to factor off both the color and singular integral factors in
\cref{eq:SingleSoftExchangeContribution} finding that
\begin{equation}
  \mathcal{A}_5^{(2), [i,j]\text{-soft}} = \overline{\mathbfcal{Z}}_n^{(1), [i,j]\text{-soft}} \mathcal{A}_5^{(1)}.
\end{equation}
We remark that our soft operator is defined to all orders in the dimensional
regulator, a natural consequence of using a local subtraction formalism.

\subsubsection{Collinear Contribution}\label{sec:CollinearContr}

Let us consider the collinear contribution to the amplitude. Similar to the soft
contribution, we can organize the full collinear contribution into a sum over
contributions associated to a given external momentum. That is, we write
\begin{equation}
  \mathcal{A}_5^{(2),\text{col}} = \sum_{j=1}^5 \mathcal{A}_5^{(2),j\text{-col}}, \qquad \text{where} \qquad \mathcal{A}_5^{(2),j\text{-col}} = \mathcal{C}_j^{(+)}[\mathcal{A}^{(2)}_5].
\end{equation}
We construct the counterterms by applying the prescription of \cref{sec:SingleIRCounterterms} to all of the integrals in \cref{eq:AllPlusFivePointAmplitude}. Notably, all factorizable contributions are collinear finite, so we discuss only the non-factorizable contributions. 
Our approach is to construct the $j\text{th}$ collinear contribution for the full amplitude and express it in a basis of integrals. This basis of integrals we could then express explicitly in terms of special functions, using, for example, $\texttt{HyperInt}$~\cite{Panzer:2014caa}.
Remarkably, we will see that explicit integration of the collinear integrals will not be necessary due to highly non-trivial cancellations.
For simplicity of exposition, we focus on the maximal topologies as the others follow from pinching. 

Let us start by considering  contributions to the planar graphs. We note that we must compute all collinear contributions to a given integral topology, as they contribute to the $j\text{th}$ collinear contribution through the permutation sum in \cref{eq:AllPlusFivePointAmplitude}. For the pentabox case the collinear contributions read
\begin{align} 
  \begin{split}
  \overline{\mathcal{C}}_{1}^{(+)} \left(\!\!
  \eqnDiag{\includegraphics[scale=0.4]{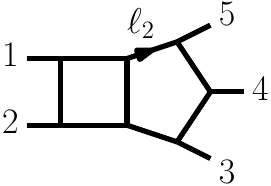}}[\mu_{22}^2]
    \right) \!=\,& \frac{1}{s_{12} \epsilon} \int_0^1 \mathrm{d} x \left(\left[\frac{1}{1-x}\right]_+ \eqnDiag{\includegraphics[scale=0.3]{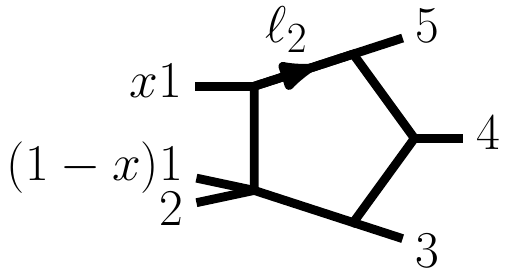}}[\mu_{22}^2] \! \right),
  \end{split}
  \label{eq:collinearPentabox1}
  \\
    \begin{split}
  \overline{\mathcal{C}}_{5}^{(+)} \left(\!\!
  \eqnDiag{\includegraphics[scale=0.4]{graphics/scalarPentabox.pdf}}[\mu_{11}^2]
    \right) \!=\,& \frac{1}{s_{34} s_{45} \epsilon} \int_0^1 \mathrm{d} x \left(\left[\frac{1}{1-x}\right]_+\eqnDiag{\includegraphics[scale=0.3]{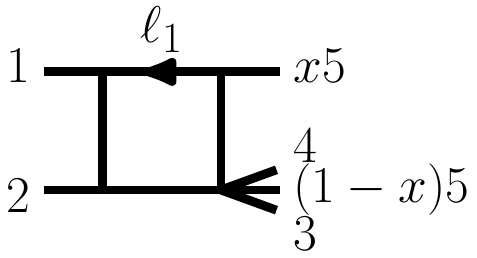}}[\mu_{11}^2]\right)- \\ & \frac{s_{35}+s_{45}}{s_{34}^2 s_{45} \epsilon} \int_0^1 \mathrm{d} x \left(\frac{1}{1-(\frac{s_{35}+s_{45}}{s_{34}})x} \eqnDiag{\includegraphics[scale=0.3]{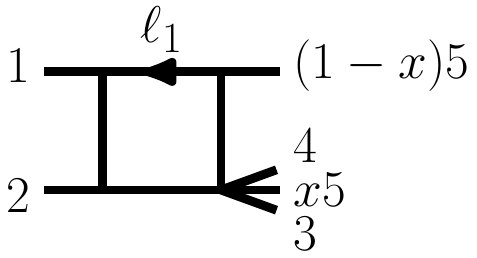}}[\mu_{11}^2] \!\right),
  \end{split}
  \label{eq:collinearPentabox5}
 \\
    \begin{split}
  \overline{\mathcal{C}}_{4}^{(+)} \left(\!\!
  \eqnDiag{\includegraphics[scale=0.4]{graphics/scalarPentabox.pdf}}[\mu_{11}^2]
    \right) \!=\,& \frac{1}{s_{34} s_{45}\epsilon} \int_0^1 \mathrm{d} x \left(\left[\frac{1}{1-x}+\frac{1}{x}\right]_+ \eqnDiag{\includegraphics[scale=0.3]{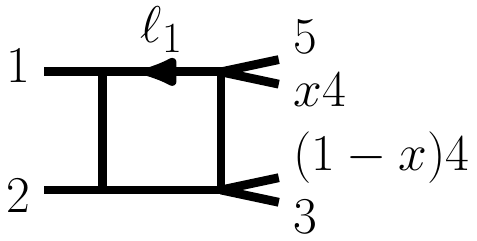}}[\mu_{11}^2] \! \right),
  \end{split}
  \label{eq:collinearPentabox4}
\end{align}
where we have performed partial fractions on the explicit $x$ dependence as well as relabelling.
As expected, the collinear contributions can be written as one-fold integrals over one-loop integrals that explicitly depend on the collinearity fraction $x$. The explicit $x$ dependence then arises from evaluating hard propagators in the singular loop on the collinear configuration.

In order to put our expression for the collinear contributions into a basis, we make use of standard integrand-level identities, cancelling numerator against denominator and lining up loop-momentum and collinearity labellings between various contributions.
This allows us to express the collinear contribution as a combination of terms that are permutations of the topologies present in~\cref{eq:collinearPentabox1,eq:collinearPentabox5,eq:collinearPentabox4}. 
Nevertheless, it turns out that, due to the collinear kinematics, there exist extra integrand identities which allow us to further reduce the set of integrals that arise. To exemplify this, we focus on constructing a relation between collinear pentagons. To this end, we consider an auxiliary hexagon topology with inverse propagators $D_i$, defined as
\begin{equation}
\begin{minipage}{0.22\textwidth}
$\eqnDiag{\includegraphics[scale=0.3]{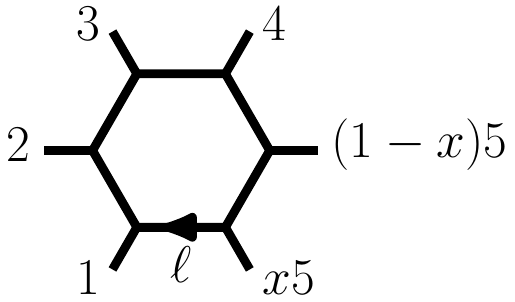}},$
\end{minipage}
\begin{minipage}{0.65\textwidth}
\vspace{-5mm}
\begin{equation*}
\begin{split}
    D_1 &= \ell^2, \hspace{21mm} D_2 = (\ell-p_1)^2, \qquad D_3=(\ell-p_{12})^2, \\
     D_4 &= (\ell+p_{45})^2, \hspace{8mm} D_5 = (\ell+p_5)^2, \qquad D_6=(\ell+x~p_5)^2.
\end{split}
\end{equation*}
\end{minipage}
\end{equation}
We can see that there is only one propagator which is $x$ dependent, as the two $x$-dependent legs are next to each other. Importantly, this hexagon integral depends only on 4 independent external momenta. 
This gives rise to an extra relation, not present for a hexagon integral arising in a genuine six-point process. To see this, we observe that we can construct a non-trivial combination of propagators that vanishes as
\begin{equation}
    \left[\frac{1}{x}+\frac{1}{1-x}\right] D_6 - \frac{D_5}{1-x}-\frac{D_1}{x} = 0.
\end{equation}
If we divide this relation through by all six inverse propagators, multiply by a numerator polynomial $N(\ell)$, and reinterpret the $D_6$ terms in terms of plus prescriptions, we find the relation
\begin{equation}\label{eq:pentgcanccol}
      \int_0^1 \mathrm{d} x \left[\frac{1}{x}\right]_+ \hspace{-2mm} \eqnDiag{\includegraphics[scale=0.3]{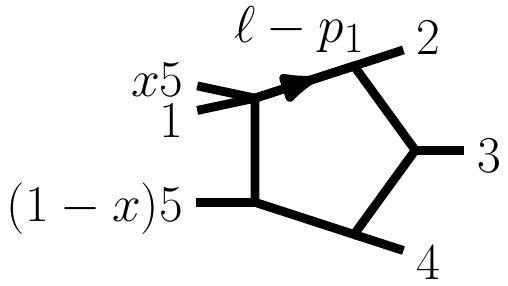}} [N(\ell)] =  \int_0^1 \mathrm{d} x \left[\frac{1}{1-x}\right]_+ \hspace{-2mm}\eqnDiag{\includegraphics[scale=0.3]{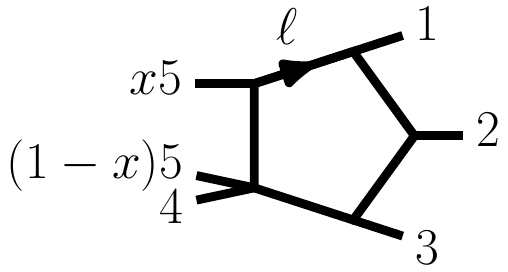}} [N(\ell)].
\end{equation}
It is important to notice that the two contributions arise from different Feynman integrals, allowing for cross-integral cancellation of the collinear contribution. We stress this class of relations is a general feature of collinear kinematics and we expect such relations to be relevant for future calculation within this formalism.

Let us now discuss the non-planar contributions and consider the reduction procedure here. We begin with the hexa-box collinear contributions, which can be written as
\begin{align}
 \begin{split}
  \overline{\mathcal{C}}_{1}^{(+)}\left(
  \!\!
  \eqnDiag{\includegraphics[scale=0.3]{graphics/hexabox.pdf}}[\mu_{22}]
    \right) = \,\,&  \frac{1}{s_{34} s_{45}\epsilon} \int_0^1 \mathrm{d} x \left(\left[\frac{1}{1-x}\right]_+\eqnDiag{\includegraphics[scale=0.3]{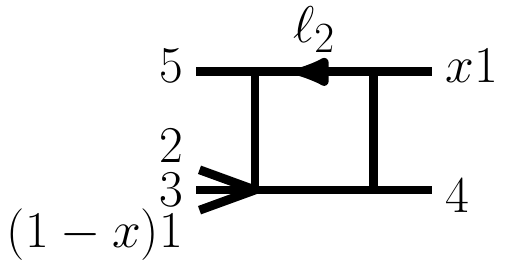}}[\mu_{22}^2]\right)- \\ & \frac{s_{35}\!+\!s_{45}}{s_{34}^2 s_{45} \epsilon} \int_0^1 \mathrm{d} x \left(\frac{1}{1\!-\!(\frac{s_{35}+s_{45}}{s_{24}})x}  \eqnDiag{\includegraphics[scale=0.3]{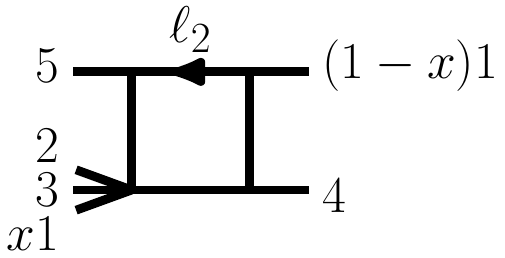}}[\mu_{22}^2]\!\right),
  \end{split}\\
   \begin{split}
  \overline{\mathcal{C}}_{2}^{(+)} \left(
  \!\!
  \eqnDiag{\includegraphics[scale=0.3]{graphics/hexabox.pdf}}[\mu_{22}]
  \!
    \right) =\,\,& \frac{1}{s_{34} s_{45}\epsilon} \int_0^1 \mathrm{d} x \left[\frac{1}{1-x}+\frac{1}{x}\right]_+ \eqnDiag{\includegraphics[scale=0.3]{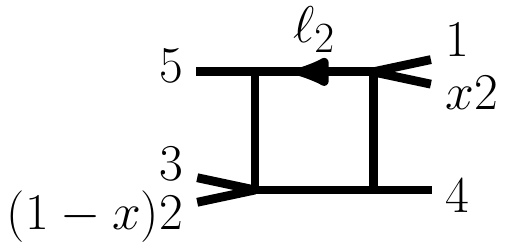}}[\mu_{22}],
  \end{split}\\
  \begin{split}
  \overline{\mathcal{C}}_{5}^{(+)} \left(
  \!\!
  \eqnDiag{\includegraphics[scale=0.3]{graphics/hexabox.pdf}}[\mu_{11}]
  \!  \right) = \,\,& \frac{1}{\epsilon} \int_0^1 \mathrm{d} x \eqnDiag{\includegraphics[scale=0.3]{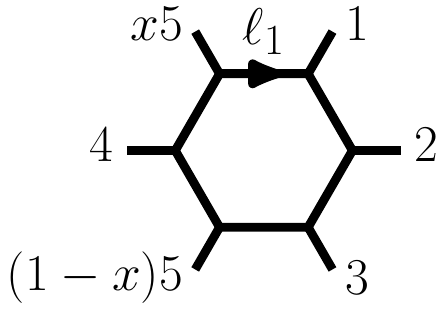}}[\mu_{11}]\,,
  \end{split}
    \label{eq:ReducibleHexagon}
\end{align}
where, again, we have performed partial fractions on the explicit $x$-dependence.
While the first two are similar to what obtained in the planar case, the contribution in \cref{eq:ReducibleHexagon} presents a new feature. 
Similarly to before, as the hexagon is an $n+1$ topology in an $n$ external leg scattering, we expect the presence of an extra relation. In this case the reduction comes from a non-trivial partial fractioning of the hexagon denominators.
To see this, let us label the denominators in \cref{eq:ReducibleHexagon} as
\begin{align}
\begin{split}
    \tilde{D}_1 &= \ell_1^2, \hspace{21mm}  \,\, \tilde{D}_2 = (\ell_1-p_1)^2, \hspace{19mm} \tilde{D}_3=(\ell_1-p_{12})^2, \\
     \tilde{D}_4 &= (\ell_1+p_{45})^2, \qquad \tilde{D}_5=(\ell_1+p_4 + x~p_5)^2, \qquad \tilde{D}_6=(\ell_1+x~p_5)^2.
\end{split}
\end{align}
The partial fractions relation can then be understood as a non-trivial way to express $1$ as a sum of denominators. One finds that
\begin{equation}
    1 = \frac{\tilde{D}_1}{x s_{45}}-\frac{\tilde{D}_6}{x s_{45}}+\frac{\tilde{D}_4}{(1-x)s_{45}}-\frac{\tilde{D}_5}{(1-x)s_{45}} \,.
\end{equation}
Inserting this into the numerator of \cref{eq:ReducibleHexagon} allows us to rewrite the hexagon collinear contribution as a combination of pentagons as
\begin{align}
  \begin{split}
  \overline{\mathcal{C}}_{5}^{(+)} \left(\!\!\!
  \eqnDiag{\includegraphics[scale=0.28]{graphics/hexabox.pdf}}[\mu_{11}]
    \!
    \right) &= \frac{1}{2 \epsilon}  \int_0^1 \frac{\mathrm{d} x}{s_{45}} \left[\frac{1}{x}\right]_+  \!\! \left( \!\!\eqnDiag{\includegraphics[scale=0.27]{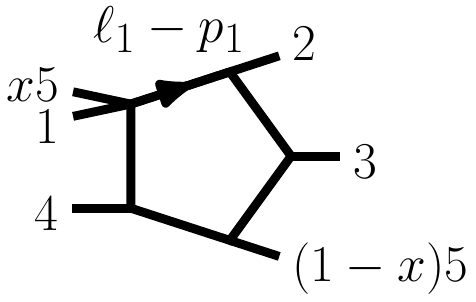}}\hspace{-4mm}[\mu_{11}]- \hspace{-2mm} \eqnDiag{\includegraphics[scale=0.27]{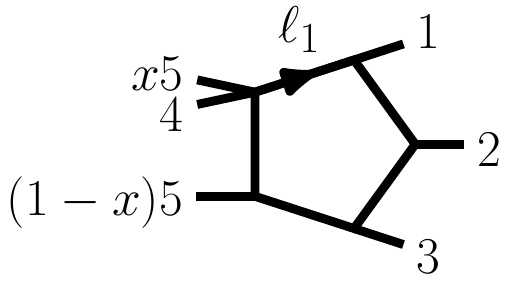}}[\mu_{11}] \!\right) + \\
    \frac{1}{2 \epsilon}  &\int_0^1 \frac{\mathrm{d} x}{s_{45}} \left[\frac{1}{1-x}\right]_+ \!\!  \left( \!\!\eqnDiag{\includegraphics[scale=0.27]{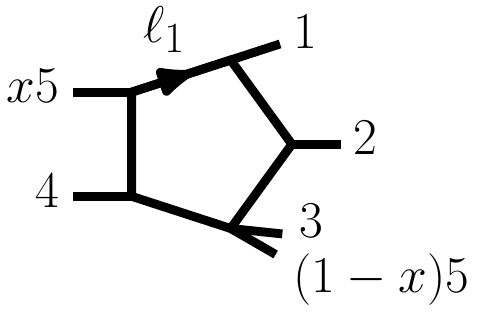}}\hspace{-4mm}[\mu_{11}]- \hspace{-2mm}\eqnDiag{\includegraphics[scale=0.27]{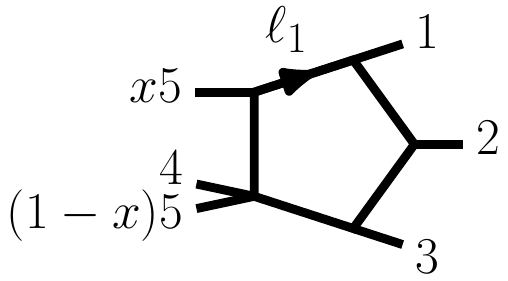}}[\mu_{11}] \! \right).
  \end{split}
\end{align}
which now reduces the integrals appearing to massive boxes and pentagons. 
It is interesting to notice that the combination coming from the partial fractioning is automatically plus-prescribed as the original hexagon integral does not have any end-point divergences.

Finally, we consider the remaining top-level, non-planar contribution: the double pentagon. In this case, the collinear contributions can be written as
\begin{align}\label{eq:collineardp}
\begin{split}
   \overline{\mathcal{C}}_{1}^{(+)} &\left(
  \eqnDiag{\includegraphics[scale=0.5]{graphics/doublePentagonBadgerLabel.pdf}}[\mu_{22}^2]
    \right) = \frac{1}{s_{12}\epsilon} \int_0^1 \mathrm{d} x \left[\frac{1}{1-x}\right]_+ \eqnDiag{\includegraphics[scale=0.3]{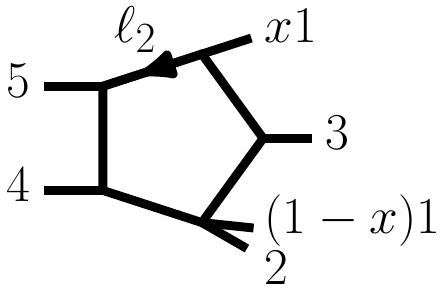}}[\mu_{22}^2]\,.
    \end{split}\\
    \begin{split}
         \overline{\mathcal{C}}_{3}^{(+)} &\left(
  \eqnDiag{\includegraphics[scale=0.5]{graphics/doublePentagonBadgerLabel.pdf}}[\mu_{22}^2]
    \right) = \frac{1}{\epsilon} \int_0^1 \mathrm{d} x \eqnDiag{\includegraphics[scale=0.3]{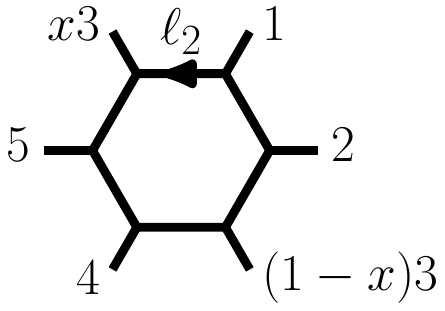}}[\mu_{22}^2]\,.
    \end{split}
\end{align}
Here, we see that the hexagon that appears has the two $x$-dependent legs separated by two external momenta. Once again we can construct a partial fractioning as in the hexa-box case and reduce it to a basis of pentagons. 

Making use of these observations, we compute all of the collinear contributions to the involved Feynman integrals and express a single collinear contribution to the amplitude in a basis of integrals. Importantly, this requires performing the non-trivial sum over color labels. In this way, we find that an individual collinear contribution to the two-loop five-point all-plus amplitude vanishes, i.e.
\begin{equation}
    \mathcal{A}_5^{(2),j\text{-col}} = 0.
    \label{eq:CollinearCancellation}
\end{equation}
This cancellation is at the level of the integrand, which greatly simplifies the computation as we do not need to perform further integrations.
We stress that the cancellation in \cref{eq:CollinearCancellation} is remarkably strong: it holds not just for the complete collinear contribution, but also for the collinear contribution to each individual leg.

\subsubsection{Ultraviolet Contribution}

Here, we compute the ultraviolet contribution to the amplitude,
$\mathcal{A}^{(2), \mathrm{UV}}_5$. This contribution naturally splits itself into
two pieces. Specifically, we have
\begin{equation}
  \mathcal{A}^{(2), \mathrm{UV}}_5 = \mathcal{A}^{(2), \mathrm{UV}, \text{non-fac}}_5 + \mathcal{A}^{(2), \mathrm{UV}, \text{fac}}_5,
\end{equation}
where the two terms on the right-hand side collect the contributions to
\cref{eq:AllPlusFivePointAmplitude} from non-factorizable and factorizable
topologies respectively.

\paragraph{Non-Factorizable Contributions}
We shall first consider the contributions from the non-factorizable topologies.
Let us begin by computing the contributions which arise from planar diagrams in
\cref{eq:AllPlusFivePointAmplitude}. In order to organize the computation, it is
useful first to construct ultraviolet counterterms, which can then be infrared
subtracted. For the ultraviolet counterterms, we find
\begin{align}
  \begin{split}
    \gamma_{\mathrm{UV}} &\left( \!\! \eqnDiag{\includegraphics[scale=0.32]{graphics/scalarPentaboxBadgerLabel.pdf}}[F_1 N_{\text{pb}}(\ell_1, \ell_2)] \right) \\
    &= (D_s - 2) \left(
      \!\!\eqnDiag{{\includegraphics[scale=0.3]{graphics/quadTadpole2.pdf}}}
      [\mu_{22}^2]
      \eqnDiag{\includegraphics[scale=0.3]{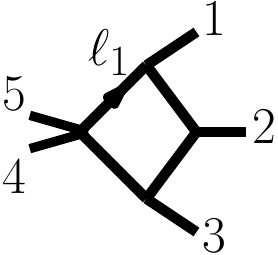}}[N_{\text{pb}}(\ell_1,
      0)] + \!\!
      \eqnDiag{\includegraphics[scale=0.3]{graphics/S45L2Triangle.pdf}}[N_{\text
        {pb}}^{\text{inf}}]
      \eqnDiag{\includegraphics[scale=0.3]{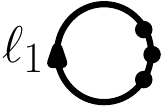}}
      [\mu_{11}^2] \right),
    \label{eq:PentaboxUV}
  \end{split}
      \\
  \gamma_{\mathrm{UV}} &\left( \!\!\eqnDiag{\includegraphics[scale=0.28]{graphics/ssdbBadgerLabel.pdf}}[F_1 N_{\text{ssdb}}] \right) 
  = (D_s - 2) \left( \!\! \eqnDiag{{\includegraphics[scale=0.3]{graphics/quadTadpole2.pdf}}} [\mu_{22}^2]
    \eqnDiag{\includegraphics[scale=0.27]{graphics/S12L1Triangle.pdf}}[N_{\text{ssdb}}]
    + \!\!
    \eqnDiag{\includegraphics[scale=0.27]{graphics/S45L2Triangle.pdf}}[N_{\text{ssdb}}]
    \eqnDiag{\includegraphics[scale=0.3]{graphics/quadTadpole1.pdf}} [\mu_{11}^2]\right) ,
   \\
  \gamma_{\mathrm{UV}} &\!\left( \!\! \eqnDiag{\includegraphics[scale=0.28]{graphics/oneMassDoubleBoxBadgerLabel.pdf}}[ F_1 N_{\text{1mdb}}] \right) \!=\! (D_s\!-\!2)\left( \!\!\! \eqnDiag{{\includegraphics[scale=0.3]{graphics/quadTadpole2.pdf}}} [\mu_{22}^2] \eqnDiag{\includegraphics[scale=0.27]{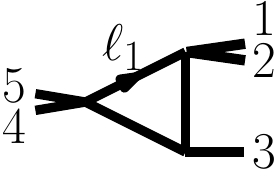}}[N_{\text{1mdb}}] \!+\!\! \eqnDiag{\includegraphics[scale=0.27]{graphics/S45L2Triangle.pdf}} [N_{\text{1mdb}}] \!\eqnDiag{{\includegraphics[scale=0.3]{graphics/quadTadpole1.pdf}}} [\mu_{11}^2]  \right).
\end{align}
Here we have defined $N_{\text {pb}}^{\text{inf}}$ as the leading term of
$N_{\text{pb}}$ in the large $\ell_1$ limit, i.e.
\begin{equation}
  N_{\text{pb}} = \ell_1^2 \left[ N_{\text{pb}}^{\text{inf}}  + \mathcal{O}\left(\frac{1}{\ell_1}\right)\right].
\end{equation}
To continue, we must now compute the infrared subtraction of these contributions.
It is clear that the tadpole piece is protected in all infrared regions by the
mass. The tadpoles therefore factorize out and we can consider the genuine
one-loop diagrams.

First, we consider the infrared subtracted triangle integrals. We only need the
scalar integrals and a simple calculation gives
\begin{align}
  (1 - \gamma_{\text{IR}}) \left( \eqnDiag{\includegraphics[scale=0.27]{graphics/S12S45L1Triangle.pdf}}[s_{12} - s_{45}] \right) &= \frac{1}{2}\left[ \log^2\left(\frac{s_{12}}{M^2}\right) - \log^2\left(\frac{s_{45}}{M^2}\right)\right] + \mathcal{O}(\epsilon),
  \\
  (1 - \gamma_{\text{IR}}) \left( \eqnDiag{\includegraphics[scale=0.3]{graphics/S12L1Triangle.pdf}}[1] \right) &= 0,
\end{align}
where the result for the two-mass triangle integral is valid in its Euclidean region.
Naturally, the infrared subtracted one-mass triangle vanishes, as we have
defined the one-mass triangle to be the soft subtraction term.
The remaining box integral to consider is a tensor integral. To compute the
(infrared-subtracted) integral, we make use of integral reduction techniques.
To achieve this in the presence of infrared subtraction, we
employ the OPP integral-reduction approach~\cite{Ossola:2006us}. Importantly,
this technique commutes with infrared subtraction, as the OPP basis of total
derivatives vanishes in infrared limits.
Therefore we can freely perform such integral reductions inside the infrared
subtraction.
Performing the integral reduction on the box integral of \cref{eq:PentaboxUV} we
find
\begin{align}
  \begin{split}
    \eqnDiag{\includegraphics[scale=0.25]{graphics/S45L1Box.pdf}}[N_{\text{pb}}(\ell_1,\!
    0)]
    \!=\! &
    \eqnDiag{\includegraphics[scale=0.25]{graphics/S45L1Box.pdf}}\!\!\left[\frac{
        s_{12} s_{23}}{2}\frac{[45]^2}{{\langle 12 \rangle \langle 23 \rangle
          \langle 31 \rangle}}\right] \!-\!
    \!\eqnDiag{\includegraphics[scale=0.23]{graphics/S12S45L1Triangle.pdf}}\!\!\left[N_{\text{1mdb}}
      \!+\! \frac{s_{12} \!-\! s_{45}}{2}\frac{[45]^2}{{\langle 12 \rangle \langle
          23 \rangle \langle 31 \rangle}} \right]
    \\
    &-
    \eqnDiag{\includegraphics[scale=0.23]{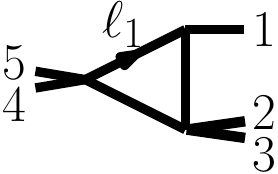}}\!\!\left[-
      N_{\text{1mdb}}|_{\substack{1\leftrightarrow 3 \\ 4 \leftrightarrow 5}} +
      \frac{s_{23} \!-\! s_{45}}{2}\frac{[45]^2}{{\langle 12 \rangle \langle 23
          \rangle \langle 31 \rangle}}\right]
    + \cdots,
  \end{split}
  \label{eq:planarBoxReduction}
\end{align}
where we suppress terms that vanish after infrared subtraction.
The last remaining ingredient is to calculate the infrared-subtracted scalar-box
integral, which is easily found to be
\begin{align}
  \begin{split}
    (1 \!-\! \gamma_{\text{IR}}) \left(\!\!\! \eqnDiag{\includegraphics[scale=0.27]{graphics/S45L1Box.pdf}}[s_{12}s_{23}] \right)
  =&\,2\left[  \mathrm{Li}_2\left( 1 \!-\! \frac{s_{12}}{s_{45}}\right) \!+\! \mathrm{Li}_2 \! \left( 1 \!-\! \frac{s_{23}}{s_{45}}\right) \!+\! \log\!\left( \frac{s_{12}}{s_{45}} \right) \log\!\left( \frac{s_{23}}{s_{45}} \right)  \!-\! \frac{\pi^2}{6} \right]
  \\
  &\quad + \frac{1}{2} \log^2\left( \frac{s_{12}}{M^2} \right) + \frac{1}{2} \log^2\left( \frac{s_{23}}{M^2} \right) - \log^2\left( \frac{s_{45}}{M^2} \right) + \mathcal{O}(\epsilon),
\end{split}
\end{align}
which we again present in the Euclidean region.

Let us now consider the contributions from the non-planar graphs in
\cref{eq:AllPlusFivePointAmplitude}. We proceed with the same strategy.
For the contributions associated to the hexabox color factor, we find that the
ultraviolet counterterms read
\begin{align}
  \gamma_{\text{UV}}\left( \!\!\!\!\eqnDiag{\includegraphics[scale=0.25]{graphics/hexabox.pdf}} [F_1 N_{\text{hb}}(\ell_1)] \right) &=(D_s-2) \eqnDiag{{\includegraphics[scale=0.3]{graphics/quadTadpole2.pdf}}}[\mu_{22}^2] \eqnDiag{\includegraphics[scale=0.25]{graphics/S45L1Box.pdf}}[N_{\text{hb}}(\ell_1)]
  \\
  \gamma_{\text{UV}}\left( \!\!\!\!\eqnDiag{\includegraphics[scale=0.25]{graphics/hexaboxpinch.pdf}} [F_1 N_{\text{1mpbx}}] \right) &= (D_s-2) \eqnDiag{{\includegraphics[scale=0.3]{graphics/quadTadpole2.pdf}}}[\mu_{22}^2] \eqnDiag{\includegraphics[scale=0.27]{graphics/S12S45L1Triangle.pdf}}[N_{\text{1mpbx}}], 
\end{align}
while for the double-pentagon color factor we find
\begin{align}
  \gamma_{\mathrm{UV}} &\left( \!\!\eqnDiag{\includegraphics[scale=0.5]{graphics/doublePentagonBadgerLabel.pdf}}[F_1 N_{\text{dp}}(\ell_1, \ell_2)] \!\right) 
  = 0,
   \\
  \gamma_{\text{UV}} &\left( \!\! \eqnDiag{\includegraphics[scale=0.35]{graphics/npdb.pdf}}\left[ F_1 N_{\text{npdb}} \right] \!\right)
  = (D_s\!-\!2) \bigg\{
    \!\!\!\!\eqnDiag{{\includegraphics[scale=0.3]{graphics/quadTadpole2.pdf}}}[\mu_{22}^2] \!\!\eqnDiag{\includegraphics[scale=0.27]{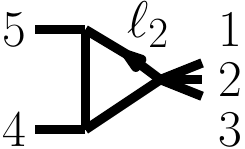}}[N_{\text{npdb}}]
    +
    \!\!\eqnDiag{{\includegraphics[scale=0.3]{graphics/quadTadpole1.pdf}}}[\mu_{11}^2] \!\eqnDiag{\includegraphics[scale=0.27]{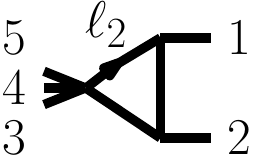}}[N_{\text{npdb}}]
                       \!\bigg\},
  \\
  \gamma_{\text{UV}}&\left( \!\!\eqnDiag{\includegraphics[scale=0.45]{graphics/pentagonBoxNP.pdf}} [F_1 N_{\text{nppb}}] \! \right)
    = (D_s\!-\!2) \eqnDiag{{\includegraphics[scale=0.3]{graphics/quadTadpole1.pdf}}}[\mu_{11}^2] \eqnDiag{\includegraphics[scale=0.27]{graphics/S45L2Triangle.pdf}}[N_{\text{nppb}}], \\
  \gamma_{\text{UV}}&\left( \!\!\eqnDiag{\includegraphics[scale=0.25]{graphics/pbx.pdf}}[F_1 N_{\text{pbx}}] \! \right)
    = (D_s\!-\!2) \eqnDiag{{\includegraphics[scale=0.3]{graphics/quadTadpole2.pdf}}}[\mu_{22}^2] \eqnDiag{\includegraphics[scale=0.27]{graphics/S12L1Triangle.pdf}}[N_{\text{pbx}}].
\end{align}
Interestingly, in this case, the double-pentagon topology has convergent
contribution has convergent ultraviolet power counting, hence there is no
ultraviolet counterterm. For the other three topologies, their ultraviolet
counterterms give one-mass triangles and hence do not survive infrared
subtraction. Non-trivially, we therefore see that the integrands associated to
the double-pentagon color factor do not give rise to any contribution to the
ultraviolet part of the amplitude.

In order to use our infrared-subtracted master integrals for the non-planar
contributions, we again must reduce a tensor box contribution to master
integrals. We find an analogous result to \cref{eq:planarBoxReduction}
\begin{align}
  \begin{split}
    \eqnDiag{\includegraphics[scale=0.25]{graphics/S45L1Box.pdf}}[N_{\text{hb}}(\ell_1,\!
    0)]
    \!=\! &
    \eqnDiag{\includegraphics[scale=0.25]{graphics/S45L1Box.pdf}}\!\!\left[\frac{
        s_{12} s_{23}}{2}\frac{[45]^2}{{\langle 12 \rangle \langle 23 \rangle
          \langle 31 \rangle}}\right] \!-\!
    \!\eqnDiag{\includegraphics[scale=0.23]{graphics/S12S45L1Triangle.pdf}}\!\!\left[N_{\text{1mpbx}}
      \!+\! \frac{s_{12} \!-\! s_{45}}{2}\frac{[45]^2}{{\langle 12 \rangle \langle
          23 \rangle \langle 31 \rangle}} \right]
    \\
    &-
    \eqnDiag{\includegraphics[scale=0.23]{graphics/S23S45L1Triangle.pdf}}\!\!\left[-
      N_{\text{1mpbx}}|_{\substack{1\leftrightarrow 3 \\ 4 \leftrightarrow 5}} +
      \frac{s_{23} \!-\! s_{45}}{2}\frac{[45]^2}{{\langle 12 \rangle \langle 23
          \rangle \langle 31 \rangle}}\right]
    + \cdots,
  \end{split}
\end{align}
where we again suppress terms which vanish after infrared subtraction.

Taking all of these ingredients together, we are able to compute the
non-factorizable contribution to the ultraviolet contribution to the amplitude.
We perform the color sum, and drop a number of terms which vanish in the
summation, finding that
\begin{equation}
  \mathcal{A}_5^{(2), \text{UV, non-fac}}
  = \kappa
  \sum_{\sigma \in S_5}
  \sigma \circ \Bigg[
  \!
  \left\{\!
    \frac{1}{2} C\!\left(\!\! \eqnDiag{\includegraphics[scale=0.26]{graphics/colorPentaboxBadgerLabel.pdf}} \right)
      \!+\!
      \frac{1}{4} C\!\left(\!\! \eqnDiag{\includegraphics[scale=0.20]{graphics/hexaboxcolor.pdf}} \! \right)
      \!
    \right\}
    \frac{[45]^2}{\langle 12 \rangle \langle 23 \rangle \langle 31 \rangle} I_{123;45}
    \Bigg],
\end{equation}
where we define
\begin{equation}
    \kappa = (D_s - 2) \eqnDiag{{\includegraphics[scale=0.3]{graphics/quadTadpole2.pdf}}} [\mu_{22}^2].
\end{equation}
and
\begin{equation}
  I_{123;45} = (1\!-\!\gamma_{\mathrm{IR}})\!
  \left(
    \!\!\! \eqnDiag{\includegraphics[scale=0.3]{graphics/S45L1Box.pdf}}\left[ \frac{s_{12} s_{23}}{2} \right] - 
  \!\eqnDiag{\includegraphics[scale=0.27]{graphics/S12S45L1Triangle.pdf}}\!\!\left[\frac{s_{12} \!-\! s_{45}}{2}\right]
  - \eqnDiag{\includegraphics[scale=0.27]{graphics/S23S45L1Triangle.pdf}}\!\!\left[\frac{s_{23} \!-\! s_{45}}{2}\right]
  \!
  \right)\!.
\end{equation}
Importantly, $I_{123;45}$ is independent of the collinear subtraction scale
$M^2$, and can be considered as the finite piece of the box integral.

\paragraph{Factorizable Contributions}
Let us now consider the computation of the contributions to the ultraviolet part
of the amplitude which arise from factorizable topologies in
\cref{eq:AllPlusFivePointAmplitude}.
Each numerator of a factorizable topology can be written as a sum
of two terms, one which is proportional to $F_2$ and the other which is
proportional to $F_3$. For the $F_2$ terms, each integral has odd degree in either
$\ell_1$ or $\ell_2$, and this momentum appears only in $\mu_{ij}$ factors.
Simple application of Passarino-Veltman reduction therefore yields that all of
the $F_2$ pieces integrate to zero and we therefore discuss
only integrals relevant to the $F_3$ piece.

To compute the factorizable (pseudo-)evanescent integrals that arise, we exploit
that both factors of the integral are themselves (pseudo-)evanescent integrals.
Let us write a factorizable $J_{\text{fac}}$ integral as $J_{\text{fac}} = J_1
J_2$ where both of $J_1$ and $J_2$ are (pseudo-)evanescent integrals of $\ell_1$
and $\ell_2$ respectively.
It is not hard to see that
\begin{equation}
  (1-\gamma_{\text{IR}})\gamma_{\text{UV}} J_{\text{fac}} =  J_{\text{fac}} + \mathcal{O}(\epsilon) = [T_{\Gamma_1} J_1][T_{\Gamma_2} J_2] + \mathcal{O}(\epsilon),
  \label{eq:FactorizablePERel}
\end{equation}
where $\Gamma_1$ and $\Gamma_2$ are the one-loop graphs associated to $J_1$ and
$J_2$ respectively.
The first relation in \cref{eq:FactorizablePERel} follows as factorizable
pseudo-evanescent integrals only have ultraviolet regions.
Therefore, there is no soft subtraction and the integral itself is equal to the
ultraviolet approximation up to $\mathcal{O}(\epsilon)$.
The second relation follows as we can then apply the counterterm logic to each
factor of $J_{\text{fac}}$, which only have ultraviolet regions.
We must therefore only calculate a collection of one-loop evanescent integrals
via the counterterm construction. The set of counterterm integrals that we
require are
\begin{align}
\begin{split}
    &T_{\Gamma_1}\left(\eqnDiag{\includegraphics[scale=0.25]{graphics/S45L1Box.pdf}}\left[ \mu_{11} \right]\right) = T_{\Gamma_1}\left(\eqnDiag{\includegraphics[scale=0.25]{graphics/S45L1Box.pdf}}\left[ \mu_{11} \ell_1^\mu \right]\right) = 0, \\
    &T_{\Gamma_2}\left(\eqnDiag{\includegraphics[scale=0.27]{graphics/S45L1Triangle.pdf}}\left[ \mu_{22} \right]\right) = \frac{6}{d-2} \eqnDiag{{\includegraphics[scale=0.3]{graphics/quadTadpole2.pdf}}}[\mu_{22}^2]\\
    &T_{\Gamma_1}\left(\eqnDiag{\includegraphics[scale=0.25]{graphics/S45L1Box.pdf}}\left[ \mu_{11} \ell_1^2 \right]\right) = \frac{10-d}{d-2} \eqnDiag{{\includegraphics[scale=0.3]{graphics/quadTadpole1.pdf}}}[\mu_{11}^2], \\
    &T_{\Gamma_1}\left(\eqnDiag{\includegraphics[scale=0.25]{graphics/S45L1Box.pdf}}\left[ \mu_{11} \overline{\ell}_1^\mu \overline{\ell}_1^\nu \right]\right) = 
     \left( \frac{6-d}{2(d-2)} g^{\mu \nu}_4 \right) \eqnDiag{{\includegraphics[scale=0.3]{graphics/quadTadpole1.pdf}}}[\mu_{11}^2], \\
    &T_{\Gamma_2}\left(\eqnDiag{\includegraphics[scale=0.27]{graphics/S45L1Triangle.pdf}}\left[ \mu_{22} \ell_2^\mu \right]\right) = \left(2 p_5^\mu + p_4^\mu \right) \frac{6-d}{d-2} \eqnDiag{{\includegraphics[scale=0.3]{graphics/quadTadpole2.pdf}}}[\mu_{22}^2],
    \end{split}
    \label{eq:OneLoopRationalIntegrals}
\end{align}
where $\Gamma_i$ are the one-loop diagrams associated to $\ell_i$ and we recall that $\overline{\ell}_1$ is the four-dimensional projection of
the loop momentum.
Correspondingly, by $g^{\mu \nu}_4$, we denote the four-dimensional restriction of the metric tensor. 
Naturally, the integrals in \cref{eq:OneLoopRationalIntegrals} agree with
well-known results for rational part integrals at one loop up to
$\mathcal{O}(\epsilon)$ corrections (see, e.g., ref~\cite{Ossola:2006us}).
All integrals that arise are either permutations of those in
\cref{eq:OneLoopRationalIntegrals}, or zero by scalelessness or
Passarino-Veltman reduction arguments.

With all of these ingredients, the computation of the ultraviolet, factorizable contribution
to the all-plus five-point amplitude is now a matter of book-keeping and
numerator algebra for one-loop integrals. We find that
\begin{equation}
  \mathcal{A}_5^{(2), \text{UV, fac}} 
  = \kappa^2
  \sum_{\sigma \in S_5}
  \sigma \circ \Bigg[ 
  \!
    \frac{1}{2} C\!\left(\!\! \eqnDiag{\includegraphics[scale=0.26]{graphics/colorPentaboxBadgerLabel.pdf}} \right) r_{\text{pb}}\!
    +
    \frac{1}{4} C \! \left( \! \! \eqnDiag{\includegraphics[scale=0.35]{graphics/doublePentagonColorBadgerLabel.pdf}}\right) r_{\text{dp}}
    \Bigg] + \mathcal{O}(\epsilon),
    \label{eq:UVNonFactorizable}
\end{equation}
where we define the rational functions
\begin{align}
\begin{split}
    r_{\text{pb}} &= 
   \frac{1}{4} \bigg(
   \frac{[13] [12]}{\langle 23\rangle  \langle 45\rangle^2}
   -\frac{5 [12]^2}{\langle 34\rangle  \langle 35\rangle  \langle
   45\rangle }+\frac{5 [14] [12]}{\langle 23\rangle  \langle 35\rangle  \langle
   45\rangle }-\frac{5 [23] [12]}{\langle 13\rangle  \langle 45\rangle
   ^2}
   -
   \frac{5 [24] [12]}{\langle 13\rangle  \langle 35\rangle  \langle
   45\rangle } +
   \\
   &\qquad \qquad\!\!\!\!
   \frac{[15] [24]}{\langle 13\rangle  \langle 23\rangle 
   \langle 45\rangle }
   -\frac{[15] [24]}{\langle 12\rangle  \langle 34\rangle  \langle
   35\rangle }
   -\frac{2 [15] [24] \langle 25\rangle }{\langle
   12\rangle  \langle 23\rangle  \langle 35\rangle  \langle 45\rangle
   }+\frac{5 \langle 12 \rangle \langle 34\rangle}{\langle 12\rangle  \langle
   35\rangle  \langle 45\rangle } +
   \\
   &\qquad\qquad\!\!\!\!
 \frac{2  \langle 35\rangle ([15] [23]-[12] [35])}{\langle 13\rangle 
   \langle 23\rangle  \langle 45\rangle ^2}
    - \frac{5 [15] [23]-3 [13]
   [25]}{\langle 12\rangle  \langle 34\rangle  \langle 45\rangle }
      +
   \frac{2 [23] [35] \langle 25 \rangle }{\langle 12\rangle ^2 \langle 45\rangle ^2} +
   \\
   &\qquad \qquad\!\!\!\!
   \frac{8 [15] [34]-11 [14] [35]}{\langle 12\rangle  \langle 23\rangle 
   \langle 45\rangle }
         -\frac{2 (4 [23] [45] + [24] [35])}{\langle 12\rangle  \langle 13\rangle  \langle 45\rangle }
   - \frac{2 [15]
   [23] \langle 25\rangle +[13] [35] \langle 35\rangle }{\langle
   12\rangle  \langle 23\rangle  \langle 45\rangle ^2} +
    \\
   &\qquad \qquad\!\!\!\!
    \frac{ \langle 25\rangle ([25][34]-[23] [45]) }{\langle 12\rangle ^2 \langle 35\rangle  \langle 45\rangle }
      -\frac{[35]
   ([24] \langle 24\rangle +2 [25] \langle 25\rangle )}{\langle
   12\rangle ^2 \langle 34\rangle  \langle 45\rangle }
    +
    \frac{4 [34] [45]}{\langle 12\rangle^2 \langle 35\rangle }
    -\frac{2 [35] [45]}{\langle 12\rangle ^2\langle 34\rangle }
   \bigg), 
   \end{split}
   \\
    r_{\text{dp}} &= \frac{3}{2}\left( 
        \frac{\langle 42 \rangle [23] [54]}{\langle 12\rangle^2 \langle 34 \rangle \langle 45 \rangle}
        -
        \frac{\langle 15 \rangle [13] [45]}{\langle 12\rangle^2 \langle 35 \rangle \langle 45 \rangle}
    \right).
\end{align}
Note that these rational functions are exactly the results of integrating the factorized terms in \cref{eq:AllPlusFivePointAmplitude}. The complexity of the expressions and the presence of non-physical poles is a consequence of the form of the integrand. Further simplifications of $\mathcal{A}_5^{(2), \text{UV, fac}} $ are possible, taking into account cancellations that occur over the color sum in \cref{eq:UVNonFactorizable}, but we do not expend any effort in uncovering them.

\subsection{Finite Remainder}

In order to understand the results of our calculation, it turns out to be
fruitful to consider them in the context of the finite remainder of the
amplitude. In the following, we define the finite remainder and explicitly
write it in terms of the region contributions to the amplitude.

\paragraph{Renormalization and infrared Factorization}

Up to the next-to-leading order (all that is necessary for two-loop amplitudes that vanish at tree level), the bare coupling is related to the renormalized
coupling $\alpha_s$ through
\begin{equation}
  \alpha_0 = \alpha_s\mu^{2\epsilon} \left( 1 - \frac{\beta_0}{\epsilon} \left( \frac{\alpha_s}{2\pi} \right) + \mathcal{O}\left(\alpha_s\right)^2\right), \qquad \text{where} \quad \beta_0 = \frac{11 C_A}{6},
\end{equation}
where $\mu^2$ is the renormalization scale (not to be confused with the
$\mu_{ij}$ that we frequently use) and the $C_A$ is the adjoint Casimir.

Once the amplitude is expressed in terms of the renormalized coupling
$\alpha_s$, the infrared divergences factorize as
\begin{equation}
  \mathcal{A}_n = {\mathbfcal{Z}}_n(\alpha_s, \epsilon) \mathcal{H}_n(\alpha_s, \epsilon).
  \label{eq:IRFactorization}
\end{equation}
Here, $\mathcal{H}_n$ is the so-called ``hard function'' that we perturbatively
expand in the renormalized coupling as
\begin{equation}
  \mathcal{H}_n(\alpha_s, \epsilon) = \sum_{l=0}^\infty \left( \frac{\alpha_s}{2\pi} \right)^l \mathcal{H}_n^{(l)}.
\end{equation}
In \cref{eq:IRFactorization}, the infrared divergences are described in terms of a single object, with
simple structure.
Specifically, one can write that
\begin{equation}
  {\mathbfcal{Z}}_n(\alpha_s, \epsilon) = \mathbb{P} \exp\left[
  \int_\mu^\infty \mathrm{d}\log(\mu') {\bf \Gamma}_n(\mu')\right] =
\sum_{l=0}^\infty \left( \frac{\alpha_s}{2\pi} \right)^l {\mathbfcal{Z}}_n^{(l)}(\epsilon),
\end{equation}
where $\mathbb{P}$ is the path-ordering symbol and ${\bf \Gamma}_n$ is the
$n$-point soft anomalous dimension matrix. Up to two-loops, this takes a
``color dipole'' form, given by
\begin{equation}
  {\bf \Gamma}_n = \sum_{1 \le i < j \le n} {{\bf T}_i \cdot {\bf T}_j} \gamma^{\mathrm{cusp}}(\alpha_s) \log \left( \frac{\mu^2}{-s_{ij} - i \varepsilon} \right) + \sum_{i=1}^n \gamma^i(\alpha_s)+ \mathcal{O}(\alpha_s)^3,
\end{equation}
where $\gamma^{\mathrm{cusp}}$ is the cusp anomalous dimension and $\gamma^{i}$
is the anomalous dimension of the $i$-th external particle, which depends on its
spin. Naturally, these have perturbative expansions,
\begin{equation}
  \gamma^{\mathrm{cusp}} = \sum_{n = 0}^{\infty} \gamma_n^{\mathrm{cusp}} \left(\frac{\alpha_s}{2\pi}\right)^{n+1}, \qquad \qquad \gamma^{i} = \sum_{n=0}^\infty \gamma_n^i \left(\frac{\alpha_s}{2 \pi}\right)^{n+1}.
\end{equation}
For the gluonic all plus case that we treat, we will need only the leading
order in these expansions, (see, e.g., refs~\cite{Korchemsky:1991zp, Harlander:2000mg})
\begin{equation}
    \gamma_0^{\mathrm{cusp}} = 2, \qquad \qquad \gamma_0^g = - \beta_0.
\end{equation}

As the all-plus amplitude vanishes at tree-level, we only need to compute
$\mathbfcal{Z}_n$ up to the one-loop level. Inserting all of the ingredients we have
collected, we find that
\begin{equation}
  \mathbfcal{Z}_n = 1 + \left(\frac{\alpha_s}{2\pi} \right) \left(  \mathbfcal{Z}_n^{(1), \mathrm{soft}} + \mathbfcal{Z}_n^{(1), \mathrm{col}} \right) + \mathcal{O}(\alpha_s^2), \\
  \label{eq:InfraredFactor}
\end{equation}
where we define
\begin{equation}
  \mathbfcal{Z}_n^{(1), \mathrm{col}} = \sum_{i} \frac{\gamma_0^i}{2 \epsilon}, \qquad
  \mathbfcal{Z}_n^{(1), \mathrm{soft}} = \sum_{1 \le i < j \le n} \mathbfcal{Z}_n^{(1), [i,j]\text{-soft}},
\end{equation}
and we abbreviate
\begin{equation}
  \mathbfcal{Z}_n^{(1), [i,j]\text{-soft}} = \gamma_0^{\mathrm{cusp}} \frac{{\bf T}_i \cdot {\bf T}_j}{2} \left[ \frac{1}{\epsilon^2} + \frac{1}{\epsilon} \log\left( \frac{\mu^2}{- s_{ij} - i \varepsilon}\right)\right] .
  \label{eq:SoftExchangeOperator}
\end{equation}

As our local subtraction is phrased at the level of the amplitude when expressed
in terms of the bare coupling, it is informative to express the amplitude in
terms of the hard function and the bare coupling. Specializing to the all-plus
case, where $\mathcal{A}_n^{(0)} = 0$, we find that
\begin{equation}
  \mathcal{A}_n = \left(  \frac{\alpha_0}{\mu^{2\epsilon}}\right)^{n/2} \! \left( \mathcal{H}_n^{(1)} + \left( \frac{\alpha_0}{2\pi} \right) \mu^{-2 \epsilon} \! \left[ \left( \mathbfcal{Z}_n^{(1), \mathrm{soft}} \!+\! \mathbfcal{Z}_n^{(1), \mathrm{col}} \!+\! n \frac{\beta_0}{2 \epsilon} \right)\mathcal{H}_n^{(1)} + \mathcal{H}_n^{(2)} \right] + \mathcal{O}(\alpha_0)^2 \right).
  \label{eq:BareDivergences}
\end{equation}
An important observation is that the
collinear anomalous dimension $\gamma_1^g$ matches the $\beta$-function, but
with opposite sign. Therefore, the ultraviolet and collinear poles cancel each other
in \cref{eq:BareDivergences}. That is,
\begin{equation}
  \mathbfcal{Z}_n^{(1), \mathrm{col}} + n \frac{\beta_0}{2\epsilon} = 0.
\end{equation}
We therefore finally find that the two-loop all-plus amplitudes divergences are
entirely controlled by the soft factor, that is
\begin{equation}
  \mathcal{A}_n^{(2)} = \mu^{-2 \epsilon} \mathbfcal{Z}_n^{(1), \mathrm{soft}} \mathcal{A}_n^{(1)} + \mathcal{H}_n^{(2)} + \mathcal{O}(\epsilon).
  \label{eq:BareDivergencesSimplified}
\end{equation}
To match the diagrammatic notation for our amplitude calculation, it is
informative to color expand the pole contribution to this formula. We
making use of the explicit form of the soft operator in
\cref{eq:SoftExchangeOperator} as well as the color decomposition of the
one-loop amplitude in
\cref{eq:OneLoopColourDecomposition}. After some algebra, one finds the simple
result that
\begin{equation}
  \mathbfcal{Z}_5^{(1), \mathrm{soft}} \!\mathcal{A}_5^{(1)} \!= \!\!\! \sum_{\sigma \in S_5 / Z_2} \!\!\! \sigma \! \circ \! \left[ C\!\left(\!\!\! \eqnDiag{\includegraphics[scale=0.32]{graphics/pentaboxColor}} \!\! \right) \! S_{12} A_5^{(1)}(1,\!2,\!3,\!4,\!5) + C\!\left(\!\!\eqnDiag{\includegraphics[scale=0.4]{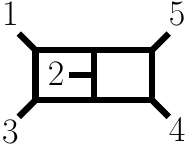}} \right) \! S_{13} A_5^{(1)}(1,\!2,\!3,\!4,\!5) \right]\!,
  \label{eq:SoftDivergencesColorDecomposition}
\end{equation}
where, for simplicity, we have defined
\begin{equation}
    S_{ij} = \frac{1}{2} \gamma_0^{\mathrm{cusp}} \left[ \frac{1}{\epsilon^2} + \frac{1}{\epsilon} \log \left(\frac{\mu^2}{- s_{ij} - i \epsilon}\right) \right].
\end{equation}
We can see that in \cref{eq:SoftDivergencesColorDecomposition}, for each color
diagram, the sub-pentagon on the right-hand-side of the diagram corresponds to
the associated one-loop amplitude. One can interpret color diagrams as
corresponding to the emission and reabsorption of a soft gluon between legs 1
and 2, or legs 1 and 3.
It is clear that analogous formulae can be written for higher multiplicity, or
for different soft exchanges.

\paragraph{Structure of Finite Remainder}

Let us now gather our ingredients and use our results to calculate the 
hard function through \cref{eq:SoftDivergencesColorDecomposition}. Due to the
structure of our ingredients, this is a remarkably simple task. Most obviously,
the collinear contribution is zero.
Next, as the ultraviolet piece is (by construction) infrared finite, it must
only be able to contribute to the hard function $\mathcal{H}_5^{(2)}$. Therefore
all of the divergences must come from the soft contribution. 
Comparing \cref{eq:SingleSoftExchangeContribution} and
\cref{eq:SoftDivergencesColorDecomposition} we find that
\begin{equation}
  \mathcal{H}_5^{(2)} = \mathcal{A}_5^{(2), \text{UV}} + \sum_{1 \le i < j \le n} \left(\overline{\mathbfcal{Z}}_5^{(1), [i,j]\text{-soft}} - \mu^{-2 \epsilon} \mathbfcal{Z}_5^{(1), [i,j]\text{-soft}}\right) \mathcal{A}^{(1)}_5 + \mathcal{O}(\epsilon).
\end{equation}
To see that the summand is finite, we note that the all-plus amplitude is finite
at one-loop, the fact that $\gamma_0^{\text{cusp}}=2$ and that
$S_{ij}$ and the triangle integral are related as
\begin{equation}
  \eqnDiag{\includegraphics[scale=0.50]{graphics/SijL1Triangle.pdf}}[s_{ij}] = \mu^{-2 \epsilon} S_{ij} + \frac{1}{2}\left[ \log^2\left( \frac{\mu^2}{-s_{ij}\!-\! i \varepsilon} \right) + \frac{\pi^2}{6}\right]+ \mathcal{O}(\epsilon).
  \label{eq:IRSchemeComparison}
\end{equation}
In summary, we find that, up to scheme dependent logarithms and $\pi^2$
contributions, the finite remainder is entirely given by the ultraviolet
contribution to the amplitude. This remarkable fact mirrors that what is
found at one-loop, where the amplitude is entirely rational.\footnote{Amusingly,
instead of expressing the amplitude in terms of $(D_s-2)$, as we find natural,
the results of~\cite{Badger:2019djh} are expressed in terms of $\kappa =
\frac{D_s-2}{6}$. Given our result, we can attribute this factor of
$\frac{1}{6}$ to the tadpole integral in our representation.}

\section{Summary and Outlook}

In this work, we have developed a new approach to the calculation of so-called ``pseudo-evanescent''
Feynman integrals: dimensionally-regulated integrals whose integrands vanish on the locus
of four-dimensional loop-momenta. To this end, we applied the local subtraction
formalism of ref.~\cite{Anastasiou:2018rib} to pseudo-evanescent integrals, which allowed us to
compute them up to and including the finite part in $\epsilon$.
Our approach naturally breaks a pseudo-evanescent
integral up into contributions which localize onto soft, collinear and ultraviolet
configurations of loop momentum space. Importantly, soft and ultraviolet
contributions are expressed as products of one-loop integrals, while collinear contributions are expressed as one-fold
integrals over one-loop integrals.
In order to demonstrate the power of our technology, we made use of the
observation that, at two loops, the special class of all-plus amplitudes can be written as a
linear combination of pseudo-evanescent integrals. We used our approach to recompute the
two-loop five-gluon all-plus amplitude, taking special care to understand the
physical origin of the contributions. We organized the all-plus amplitude
into contributions from soft, collinear and ultraviolet regions. Strikingly, we observe
that, region by region, the soft and collinear contributions sum to exactly the universal
expectation of the infrared pole structure. The remaining contributions are
of ultraviolet origin and contribute only to the finite remainder. 

Now that a technology exists that allows a simple computation of pseudo-evanescent
integrals beyond the one-loop level, we foresee a number of interesting next steps. 
Firstly, it would be interesting to systematically understand the decomposition of the amplitude 
integrand into four-dimensional and pseudo-evanescent parts. 
Given that we can compute pseudo-evanescent integrals in a streamlined manner, the 
problem is simplified to computing the four-dimensional piece. 
This potentially provides important simplifications, such as the use of
four-dimensional integrand construction methods.
Secondly, an important technical question that arises when constructing such a
formalism is to ensure that the use of integration by parts identities does
not mix the pseudo-evanescent and four-dimensional decomposition. 
We expect that techniques similar to those of ref.~\cite{DeAngelis:2025agn} will 
be useful in studying this question.
Finally, it is intriguing that, in the five-point two-loop all-plus
amplitude, we find that all soft and collinear pseudo-evanescent contributions cancel completely against universal infrared behavior, including the finite piece. 
It would be interesting to understand if this phenomenon holds more generally and if this can be broadly exploited in Standard Model amplitude calculations.

\section*{Acknowledgements}
We would like to thank S.~Caron-Huot, Z.~C. Chan, H.~Ita, D.~Kosower,
P.~Novichkov and V.~Sotnikov for helpful discussions. A.G. would like to thank
B.~Basso for useful discussions and encouragement.
The work of B.~P. received support from the French Agence Nationale pour la
Recherche, under grant ANR--17--CE31--0001--01.
This work was supported by the French Agence Nationale pour la Recherche,
under grant ANR--17--CE31--0001--02.
This project has received funding from the European’s Union Horizon 2020
Research and Innovation Programme under grant agreement number 896690, project
‘LoopAnsatz’. A.~G.~is supported by a Royal Society funding, URF\textbackslash{R}\textbackslash221015. A.~G. would also like to thank Nordforsk for partial support.

\newpage

\appendix

\crefalias{section}{appendix}
\section{Collinear Kernel}
\label{app:CollinearKernelDerivation}

It is and instructive exercise to derive \cref{eq:CollinearCounterterm}. The first term is a scaleless integral which
vanishes. We next write the whole counterterm as 
\begin{equation}
  \frac{1}{i \pi ^{D/2} }\int \mathrm{d}^{D}\ell_i \frac{G(x_j[\ell_i]p_j)}{(\ell_i^2 - M^2)([\ell_i-p_j]^2 - M^2)}
  = \int_{-\infty}^{\infty} \mathrm{d}x I_j(x) G(x p_j),
  \label{eq:CollinearKernelFirstStep}
\end{equation}
where we introduce the collinear kernel
\begin{equation}
  I_j(x) = \frac{1 }{i \pi^{D/2}} \int \mathrm{d}^D \ell_i \frac{\delta\left(x - \frac{\ell_i \cdot \eta_j}{p_j \cdot \eta_j}\right)}{(\ell_i^2 - M^2)([\ell_i - p_j]^2 - M^2)}.
\end{equation}
The task is then to compute $I_j(x)$.
The first step is to make a change of integration variables given by
\begin{equation}
\ell_i = x_j p_j + \beta_j \frac{\eta_j}{2 \eta_j\cdot p_j} + \ell_{i, \bot},
\end{equation}
where $\eta_j^2 = p_j \cdot \ell_{i, \bot} = \eta_j \cdot \ell_{i, \bot} = 0$.
Due to the judicious normalization of the $\beta_j$ term, the Jacobian is simply
$\frac{1}{2}$. The region of integration is unchanged.
The two propagators are given by 
\begin{equation}
  \ell_i^2 = \ell_{i, \perp}^2 + x_j \beta_j, \qquad (\ell_i-p_j)^2 =  \ell_{i, \perp}^2 + (x_j-1) \beta_j.
\end{equation}
Performing the integral over the $\delta$-function, one finds
\begin{equation}
  I_j(x) = \frac{1 }{2 i \pi^{D/2}} \int \mathrm{d} \beta_j \mathrm{d}^{D-2} \ell_{i, \perp} \frac{1}{(\ell_{i, \perp}^2 + x_j \beta_j - M^2 + i \epsilon)(\ell_{i, \perp}^2 + [x_j-1] \beta_j - M^2 + i \epsilon)},
\end{equation}
where we have explicitly reintroduced the causal $i \epsilon$ prescription.

One can now compute the $\beta_j$ integral via the residue theorem. The two
poles in $\beta_j$ are given by
\begin{equation}
  \beta_j^{(0)} = \frac{1}{x}(M^2 - \ell_{i, \perp}^2) - i\epsilon \, \mathrm{Sgn}(x),
  \qquad
  \beta_j^{(1)} = \frac{1}{x-1}(M^2 - \ell_{i, \perp}^2) - i\epsilon \, \mathrm{Sgn}(x-1).
\end{equation}
For the integral to be non-zero, we must have two poles on opposite sides of the
real line, hence we are restricted to the range $0 \le x \le 1$. Closing the
contour on the upper half plane and computing the residue, one finds
\begin{equation}
  I_j(x) = -\frac{\Theta(0 \le x \le 1)}{\pi^{(D-2)/2}} \int \frac{\mathrm{d}^{D-2}\ell_{i,\perp}}{\ell_{i,\perp}^2 - M^2 + i \epsilon}.
\end{equation}
The remaining tadpole-like integral is performed over space-like configurations,
hence no Wick rotation is required and we find
\begin{equation}
  I_j(x) = \Theta(0 \le x \le 1) \frac{\Gamma( 1 + \epsilon )}{\epsilon} M^{-2\epsilon}.
\end{equation}
Inserting this into \cref{eq:CollinearKernelFirstStep}, we
find the result of \cref{eq:CollinearCounterterm}.

\section{All Plus Five-Point Numerators}
\label{sec:AllPlusFivePointIntegrand}

In this appendix, we record the full five-point two-loop integrand, first
computed in ref.~\cite{Badger:2015lda}, in the conventions used in this work.
We employ the spinor-helicity formalism, with massless external momenta
decomposed as $p_i^{\dot\alpha\alpha} = \tilde\lambda_i^{\dot\alpha}\lambda_i^\alpha$.
The angle and square spinor brackets are denoted $\langle ij \rangle$ and $[ij]$
respectively, and satisfy $\langle ij \rangle [ji] = s_{ij} = 2\,p_i \cdot p_j$.
Two commonly appearing spinorial objects are
\begin{align}
  \mathrm{tr}_5 &= [12]\langle 23 \rangle[34]\langle 41 \rangle - \langle 12 \rangle[23] \langle 34 \rangle[41], \\
  \mathrm{tr}_+(ijkl) &= [ij]\langle jk \rangle[kl]\langle li \rangle.
\end{align}
alongside the vector transverse to $p_a$, $p_b$ and $p_c$ given by
\begin{align}
  \omega_{abc}^\mu = \frac{\langle bc \rangle [ca]}{s_{ab}}\frac{\langle a|\gamma^\mu|b]}{2}
                   - \frac{\langle ac \rangle [cb]}{s_{ab}}\frac{\langle b|\gamma^\mu|a]}{2}\,,
  \label{eq:OmegaDef}
\end{align}
which enters the numerators below through the scalar products $\ell_i \cdot \omega_{abc}$.

The numerators for the non-factorizable topologies are given by
\begin{align}
  N_{\text{pb}}(\ell_1, \ell_2) &= -\frac{s_{12}s_{23}s_{45}}{\langle 12 \rangle \langle 23 \rangle \langle 34 \rangle \langle 45 \rangle \langle  51 \rangle \mathrm{tr}_5} \left( \mathrm{tr}_+(1345)(\ell_1+p_5)^2 + s_{15}s_{34}s_{45} \right), \\
  N_{\text{ssdb}} &= \frac{s_{12}s_{23}s_{34}s_{45}s_{51}}{\langle 12 \rangle \langle 23 \rangle \langle 34 \rangle \langle 45 \rangle \langle  51 \rangle \mathrm{tr}_5}, \\
  N_{\text{1mdb}} &= N_{\text{1mpbx}} = N_{\text{nppb}} =-\frac{s_{34}s_{45}^2\mathrm{tr}_+(1235)}{\langle 12 \rangle \langle 23 \rangle \langle 34 \rangle \langle 45 \rangle \langle  51 \rangle \mathrm{tr}_5}, \\
  N_{\text{npdb}} &= -\frac{s_{12} s_{45}}{4 \langle  12 \rangle \langle 23 \rangle \langle 34 \rangle \langle  45 \rangle \langle 51 \rangle \mathrm{tr}_5}\left( 2 s_{23} s_{34} s_{15} - s_{23} \mathrm{tr}_+(1345) + s_{34} \mathrm{tr}_+(1235) \right), \\
  N_{\text{pbx}} &= -\frac{s_{12}}{2 \langle  12 \rangle \langle 23 \rangle \langle 34 \rangle \langle  45 \rangle \langle 51 \rangle \mathrm{tr}_5}\left( s_{23} s_{45} \mathrm{tr}_+(1435) - s_{15} s_{34} \mathrm{tr}_+(2453) \right), \\
  \begin{split}
    N_{\text{dp}}(\ell_1, \ell_2) &= \frac{s_{12}s_{45}}{4 \langle 12 \rangle \langle 23
      \rangle \langle 34 \rangle \langle 45 \rangle \langle 51 \rangle
      \mathrm{tr}_5}
    \Big( s_{23} \mathrm{tr}_+(1345)\left( 2 s_{12} - 4 \ell_1 \!\cdot\! (p_5 \!-\! p_4) + 2 (\ell_1 \!-\! \ell_2) \!\cdot\! p_3 \right) \\
      & \hspace{48mm}- s_{34} \mathrm{tr}_+(1235)(2 s_{45} - 4 \ell_2 \!\cdot\! (p_1 \!-\! p_2) - 2 (\ell_1 \!-\! \ell_2)\!\cdot\! p_3) \\
      & \hspace{48mm}- 4 s_{23} s_{34} s_{15} (\ell_1 - \ell_2)\!\cdot\! p_3
    \Big),
  \end{split}
  \\
  N_{\text{hb}}(\ell_1) &= -\frac{s_{12}s_{23}s_{45}}{\langle 12 \rangle \langle 23 \rangle \langle 34 \rangle \langle 45 \rangle \langle 51 \rangle \mathrm{tr}_5}\left( \mathrm{tr}_+(1345)\left(\ell_1 \cdot[p_4 - p_5] - \frac{s_{45}}{2}\right) + s_{15}s_{34}s_{45} \right), 
\end{align}
while the numerators for the factorizable topologies are given by
\begin{equation}
  N_{\mathrm{bt}}(\ell_1, \ell_2) = - \frac{s_{12} \mathrm{tr}_+(1345)}{2 \langle 12 \rangle \langle 23 \rangle \langle 34 \rangle \langle 45 \rangle \langle 51 \rangle s_{13}} (2 (\ell_1 \cdot \omega_{123}) + s_{23})\left(F_2 + F_3 \frac{(\ell_1+\ell_2)^2+ s_{45}}{s_{45}}\right),
\end{equation}
\begin{equation}
  N_{\mathrm{tt1m}}(\ell_1, \ell_2) = - \frac{-(s_{45} - s_{12}) \mathrm{tr}_+(1345)}{2 \langle 12 \rangle \langle 23 \rangle \langle 34 \rangle \langle 45 \rangle \langle 51 \rangle s_{13}} (2 (\ell_1 \cdot \omega_{123}) + s_{23})\left(F_2 + F_3 \frac{(\ell_1+\ell_2)^2+ s_{45}}{s_{45}}\right), 
\end{equation}
\begin{align}
  \begin{split}
    N_{\text{sstt}}(\ell_1, \ell_2) &= -\frac{1}{\langle  12 \rangle \langle 23 \rangle \langle 34 \rangle \langle 45 \rangle \langle 51 \rangle} \times
    \\
    \Bigg\{\!\frac{1}{2}\Bigg( \! \mathrm{tr}_+(12&45)  \!-\! \frac{\mathrm{tr}_+(1345) \mathrm{tr}_+(1235)}{s_{13} s_{35}}\!\Bigg) \left(\! F_2 \!+\! F_3 \frac{4 (\ell_1 \!\cdot\! p_3)(\ell_2 \!\cdot\! p_3) \!+\! (\ell_1\!+\!\ell_2)^2(s_{12} \!+\! s_{45}) \!+\! s_{12}s_{45}}{s_{12}s_{45}} \!\right) \\
    + F_3\Bigg[
      (\ell_1 + &\ell_2)^2 s_{15} + 
      \mathrm{tr}_+(1235)\left( \frac{(\ell_1 \!+\! \ell_2)^2}{2 s_{35}} - \frac{\ell_1 \cdot p_3}{s_{12}}\left( 1 \!+\! \frac{2 (\ell_2 \cdot \omega_{543})}{s_{35}} \!+\! \frac{s_{12} \!-\! s_{45}}{s_{35} s_{45}}(\ell_2 \!-\! p_5)^2 \right) \right)
      \\
      + \mathrm{tr}_+&(1345)\left( \frac{(\ell_1 + \ell_2)^2}{2 s_{13}} - \frac{\ell_2 \cdot p_3}{s_{45}}\left( 1 + \frac{2 (\ell_1 \cdot \omega_{123})}{s_{13}} + \frac{s_{45} - s_{12}}{s_{12} s_{13}}(\ell_1 - p_1)^2 \right) \right)
    \Bigg]
    \Bigg\},
  \end{split}
                       \\
  \begin{split}
    N_{\text{npsstt}}(\ell_1, \ell_2) &= \frac{F_3}{2 \langle  12 \rangle \langle 23 \rangle \langle 34 \rangle \langle 45 \rangle \langle 51 \rangle s_{12} } \times
    \\
    &\Bigg\{
    (s_{45} - s_{12})\mathrm{tr}_+(1245) - \left( \mathrm{tr}_+(1245)  - \frac{\mathrm{tr}_+(1345) \mathrm{tr}_+(1235)}{s_{13} s_{35}} \right)2 (\ell_1 \cdot p_3)
    \\
    & - \frac{s_{45} \mathrm{tr}_+(1235)}{s_{35}}\left( 2 (\ell_2 \cdot \omega_{543}) + \frac{s_{12} - s_{45}}{s_{45}} (\ell_2 - p_5)^2 \right)
    \\
    & - \frac{s_{12} \mathrm{tr}_+(1345)}{s_{13}}\left( 2 (\ell_1 \cdot \omega_{123}) + \frac{s_{45} - s_{12}}{s_{12}} (\ell_1 - p_1)^2 \right)
    \Bigg\}.
  \end{split}
\end{align}
Here, we make use of two $\epsilon$-dimensional objects,
\begin{align}
  F_2 &= 4(D_s - 2) (\mu_{11} + \mu_{22})\mu_{12}, \\
  F_3 &= (D_s - 2)^2 \mu_{11}\mu_{22}.
\end{align}

\bibliographystyle{JHEP}
\bibliography{bibliography}

\end{document}